\documentclass[preprint,aps,prc,showpacs,nofootinbib,floatfix]{revtex4-2}
\usepackage{amsmath}

\usepackage{amssymb}
\usepackage{graphics}
\usepackage{epsfig}

\usepackage{subfigure}

\newcommand\nn{\nonumber}
\newcommand\ba{\begin{eqnarray}}
\newcommand\ea{\end{eqnarray}}
\newcommand\be{\begin{equation}}
\newcommand\ee{\end{equation}}
\begin{document}

\title{Charged pion production in the annihilation reactions $e^+ + e^- \to p + \bar{n} +\pi^-$ and $e^+ + e^- \to n + \bar{p} +\pi^+.$
Contribution of the nonresonant mechanism}


\author{M.I. Konchatnij}
\email{konchatnij@kipt.kharkov.ua}

 \affiliation{\it National Science Centre, Kharkov Institute of
Physics and Technology, Akademicheskaya 1, and V. N. Karazin
Kharkov National University, Dept. of
Physics and Technology, 31 Kurchatov, 61108 Kharkov, Ukraine}
\author{N.P. Merenkov}
\email{merenkov@@kipt.kharkov.ua}
\affiliation{\it National Science Centre, Kharkov Institute of
Physics and Technology, Akademicheskaya 1, and V. N. Karazin
Kharkov
National University, Dept. of
Physics and Technology, 31 Kurchatov, 61108 Kharkov, Ukraine}

\begin{abstract}
The nonresonant mechanism in the reactions $e^+\,e^-\to \pi^-\,p\,\bar{n}$ and $e^+\,e^-\to \pi^+\,n\,\bar{p}$ is investigated in frame of the one-photon exchange approximation. The description of the hadronic phase space and the invariant amplitude  formalism, earlier developed for the neutral pion channel, are applied here. Some modification is required to fulfill the  conservation of the hadron electromagnetic current and to account for an additional contribution due to diagram with the pion pole.  The double and single distributions over the invariant variables as well the total cross section,  are obtained. The results are plotted for $s=$5,\,10,\,16\, GeV$^2$. The calculations are performed for the
$\pi^-\,p\,\bar{n}$ channel and the rules to proceed to the
$\pi^+\,n\,\bar{p}$ channel  are formulated for every distribution.
\end{abstract}

\maketitle

\section{Introduction}
\label{Section:Introduction}

In our previous works \cite{Gakh:2022fad,Gakh:2022fpf}
we considered the annihilation reaction
\begin{equation}
e^+(k_1) + e^- (k_2) \to N(p_1) +\bar N (p_2) + \pi^0(k),
N = p, n
\label{eq:1}
\end{equation}
that is in principle accessible at BESIII \cite{Yuan:2019zfo}. In these papers we have performed the general analysis of the differential cross section and polarization
observables for neutral pion emission and obtained  double- and single distributions over the invariant variables in frame of  the so-called nonresonant mechanism.

The present analysis assumes the conservation of the hadron electromagnetic current (CHC) and the P-invariance of the hadron electromagnetic interaction. In these conditions the matrix element of the process (\ref{eq:1}) can be expressed in terms of six invariant amplitudes, and all the physical observables are expressed in terms of their bilinear combinations. This statement  holds for any dynamical mechanism including also the resonant ones.

The key moment in our numerical estimations is the choice of the nucleon (proton and neutron) electromagnetic form factors (FF's),  for a review, see \cite{Pacetti:2014jai}.
We used two parameterizations of FF's, which are based on the VMD model and perturbative QCD and were labeled in \cite{Gakh:2022fad} as "old" \cite{Iachello:1972nu,Iachello:2004aq} and "new" \cite{Bijker:2004yu} versions.

\begin{figure}
\centering
\includegraphics[width=0.322\textwidth]{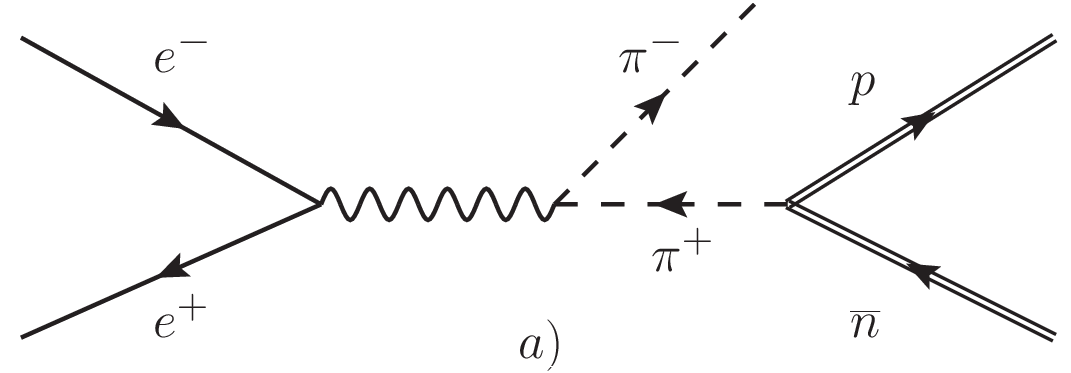}
\includegraphics[width=0.322\textwidth]{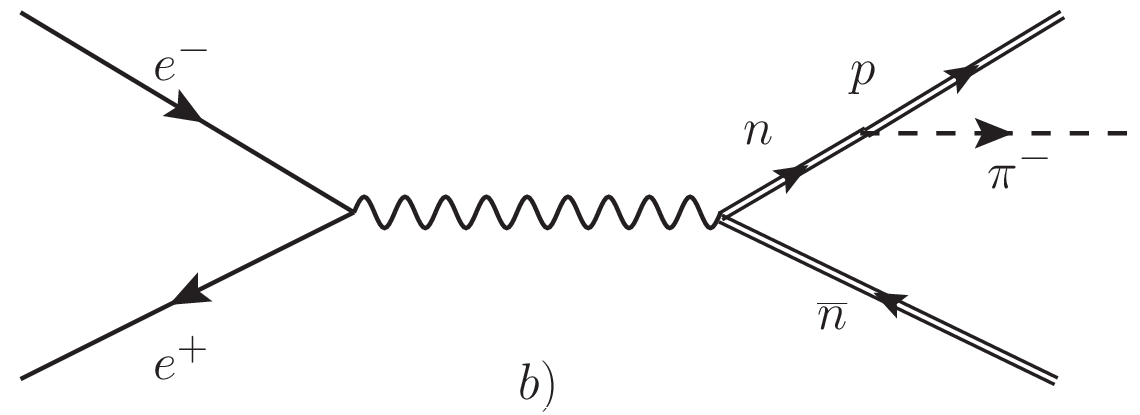}
\includegraphics[width=0.322\textwidth]{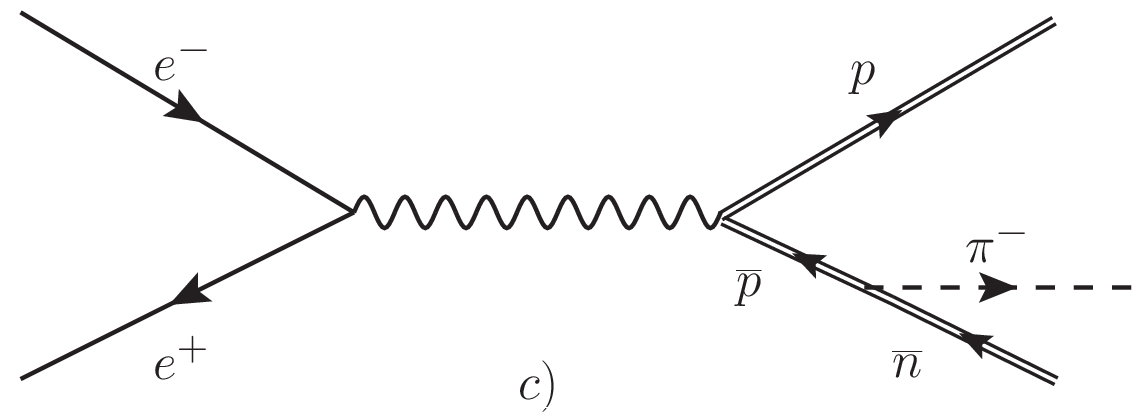}
 \parbox[t]{0.9\textwidth}{\caption{Feynman diagrams  corresponding to the nonresonant mechanism for the  reaction $e^+ + e^- \to p + \bar{n} +\pi^-$.}
\label{Fig:Fig1} }
\end{figure}

Here we apply the formalism previously developed to describe the annihilation processes  with production of a charge pion
\begin{equation}
\label{eq:2}
e^+(k_1) + e^- (k_2) \to p(p_1) +\bar{n}(p_2) + \pi^-(k), \ \ e^+(k_1) + e^-(k_2) \to n(p_1) +\bar{p}(p_2) + \pi^+(k).
\end{equation}
Let us consider, for definitness, the production of $\pi^-$-meson,
namely $e^+e^-\to \pi^- p\bar n$. There are three diagrams in which the real charged pion is connected with the $\gamma^* \pi^-\pi^+,\,\,n\pi^-\,p $ and $\bar{p}\pi^-\bar{n}$ vertices (see Fig.\,\ref{Fig:Fig1}). In the case of neutral pions, the first diagram (Fig.\, \ref{Fig:Fig1}a) does not contribute, and the electromagnetic current corresponding to sum of the two diagrams with the $\pi^0 N \bar{N},\,N=p,\,n$ interaction is conserved, but for the charged pion
 production the part of the nucleon electromagnetic current (diagrams (Fig.\,\ref{Fig:Fig1}b,c)) as well the part due to pion electromagnetic current (Fig.\, \ref{Fig:Fig1}a) are not.

In order to apply the formalism developed in Ref.  \cite{Gakh:2022fad} we have to reconstruct the CHC, what we are reaching by the well known modification of the full hadron electromagnetic current in such a way that \cite{Akhiezer:1977nd}
\begin{equation}\label{eq:I3}
J_\mu \to \bar{J}_\mu = J_\mu - \frac{J\cdot q}{q^2} q_\mu, \ \ q\cdot \bar{J} =0, \ \ q=(k_1+k_2)= (p_1+p_2+k),
\end{equation}
and to use $\bar{J}_\mu$ instead of $J_\mu$. It can be done because $(J\cdot q) =0$ at $q^2 =0,$ and the matrix elements of processes (\ref{eq:2}) do not affected by such modification due to conservation of the lepton current $(q\cdot l) = 0$. (For hadron and lepton currents see below).
In this case the pion electromagnetic form factor, $F_\pi(q^2),$ has to be known in the time-like region $q^2>0$.

\section{Formalism for the reaction $e^+ + e^- \to p + \bar{n} +\pi^-$}
\label{Section2:piminus}
Let us remind briefly a formalism of the invariant amplitudes used in \cite{Gakh:2022fad} for the description of the process $e^+ + e^- \to N + \bar{N} +\pi^0$.
The matrix element in the one photon approximation can be written as the convolution of the lepton $(e l_{\mu})$ and hadron $(e J_{\mu})$ currents
\begin{equation}\label{eq:MlJ4}
\mathcal{M}=\frac{e^2}{q^2}l^{\mu}J_{\mu}, \ \ l^{\mu}=\bar v(k_2)\gamma^{\mu}u(k_1),
\end{equation}
where $e$ is the electromagnetic coupling $\alpha = e^2/(4\pi) = 1/137$ and $k^2=m^2, \,p_1^2=p_2^2=M^2$, $m (M)$ is the pion (nucleon) mass. Further, we neglect the electron mass where it is possible.

The general form of the matrix element has been chosen in analogy with the
process of the pion electroproduction on the nucleons \cite{Adler:1968tw}:
\begin{equation}
\label{eq:M5}
\mathcal{M}=\frac{e^2}{q^2}\varphi^+_{\pi}\Sigma_{i=1}^6\bar u(p_1)\gamma_5M_i v(p_2)A_i, \ \ \gamma_5=i \gamma^0 \gamma^1 \gamma^2 \gamma^3,
\end{equation}
where the $M_i$ structures have the following form
\begin{eqnarray}
\label{eq:Mi6}
M_1&=&\frac{1}{2}\gamma^{\mu}\gamma^{\nu}F_{\mu\nu}, \ \ M_2=p^{\mu}k^{\nu}F_{\mu\nu}, \ \
M_3=\gamma^{\mu}k^{\nu}F_{\mu\nu}, \ \
M_4=(\gamma^{\mu}p^{\nu}-2M\gamma^{\mu}\gamma^{\nu})F_{\mu\nu},
\nn\\
M_5&=&q^{\mu}k^{\nu}F_{\mu\nu}, \ \
M_6=q^{\mu}\gamma^{\nu}F_{\mu\nu}, \ \
F_{\mu\nu}=l_{\mu}q_{\nu}-l_{\nu}q_{\mu},
\end{eqnarray}
and $p=p_1-p_2$. The invariant amplitudes $A_i$ (i=1-6) are complex functions of three independent variables: for example, $q^2$  and $s_{1,2}=(p_{1,2}+k)^2.$

Eqs. (\ref{eq:MlJ4},\ref{eq:M5}) imply that, in the general case,
$ J_{\mu}$ can be written as follows
\begin{equation}\label{eq:JO7}
J_{\mu}=\varphi^+_{\pi}\bar u(p_1)\gamma_5\hat O_{\mu}v(p_2),
\end{equation}
where the matrix $\hat O_{\mu}$ has the form
\begin{eqnarray}\label{eq:O8}
\hat O_{\mu}&=&(k\cdot qp_{\mu}-p\cdot qk_{\mu})A_2-q^2\tilde k_{\mu}A_5+(k\cdot qA_3+p\cdot qA_4-q^2A_6
)\gamma_{\mu}+\nn\\
&&+(A_6q_{\mu}-A_4p_{\mu}-A_3k_{\mu})\hat q+(A_1-4MA_4)(\gamma_{\mu}\hat q -q_{\mu}),
\end{eqnarray}
where $\tilde k_{\mu}=k_{\mu}-(k\cdot q/q^2)q_{\mu}$. Eq.  (\ref{eq:O8}) holds when the hadron electromagnetic current (\ref{eq:JO7}) satisfies the condition $J_\mu\,q^\mu =0$.

Our goal is to express the invariant amplitudes in terms of the nucleon and pion form factors and the scalar products of the hadron 4-momenta.

The pion current for the transition $\gamma^*\to \pi^-(k_2)\,\,\pi^+(k_1)$ can be written as
\begin{equation}\label{eq:pion}
\mathfrak{J}_{\mu}=\varphi_1^*(k_1)\varphi_1^*(k_2)F_{\pi}(q^2)(k_1-k_2)_{\mu}.
\end{equation}
Then the part of hadron current corresponding to the diagram Fig.~\ref{Fig:Fig1}a), where the virtual photon $\gamma^*(q)$ transforms into pion pair
$\pi^-(k)\,\,\pi^+(q-k)$ is
\begin{equation}\label{eq:pioncurrent}
J_{\mu}^{\pi}=\frac{g}{d_3}F_{\pi}(q^2)\bar u(p_1)\gamma_5v(p_2)(q-2k)_{\mu}, \ d_3=q^2-2q\cdot k,
\end{equation}
where $g$ is the $p\bar n\pi^-$ coupling constant $(g=\sqrt{2}\,g_{\pi^0 N N}).$

The parts of the hadron electromagnetic current corresponding to diagrams Fig.~\ref{Fig:Fig1}b),c), where $\pi^-$-meson is emitted from vertices
$\bar n\to \pi^- p$ and $\bar p\to \pi^- \bar n$, respectively, are
\begin{equation}\label{eq:neuteroncurrent}
J_{\mu}^n=\frac{g}{d_2}\bar u(p_1)\gamma_5(\hat q-\hat p_2 +M)\Gamma_{\mu}^n(q)v(p_2), \ d_2=q^2-2q\cdot p_2,
\end{equation}
\begin{equation}\label{eq:protoncurrent}
J_{\mu}^p=\frac{g}{d_1}\bar u(p_1)\Gamma_{\mu}^p(q)(\hat p_1-\hat q +M)\gamma_5v(p_2), \ d_1=q^2-2q\cdot p_1,
\end{equation}
where
\begin{equation}\label{eq:FF3}
\Gamma_{\mu}^i(q)=F_{1}^i(q^2)\gamma_{\mu}+\frac{F_{2}^i(q^2)}{4M}(\hat q\gamma_{\mu}-\gamma_{\mu}\hat q), \ i=p,\,n.
\end{equation}

Taking into account that
\begin{equation}\label{eq:totalgauge}
q^{\mu}(J_{\mu}^p+J_{\mu}^n+J_{\mu}^{\pi})=gR[F_{\pi}(q^2)-(F_{1}^p(q^2)-F_{1}^n(q^2))], \ R=\bar u(p_1)\gamma_5v(p_2),
\end{equation}
we have
\begin{equation}\label{eq:barJ}
\bar{J}_\mu=J_\mu^p +J_{\mu}^n + J_\mu^\pi -\frac{q_\mu}{q^2}\,g\,R\big[F^{\pi} - F_1^p + F_1^n\big].
\end{equation}
Because of $F_1^p(0) = F_\pi(0) =1,\ F_1^n(0) =0,$  at $q^2=0$ we have $(q\cdot J(0)) =0,$ as it was noted above.

If we use $\bar{J}_\mu$ as defined by relation (\ref{eq:barJ}), the invariant amplitudes in (\ref{eq:O8}) are
\begin{eqnarray}
 A_1 &=& g\bigg[\frac{F_2^p - F_1^p}{k\cdot q-p\cdot q}+ \frac{F_2^n -F_1^n}{k\cdot q+p\cdot q}\bigg]\,, \ \ A_2= \frac{g}{k\cdot q}\bigg[\frac{F_1^p}{k\cdot q-p\cdot q}+ \frac{F_1^n}{k\cdot q+p\cdot q}\bigg]\,,\nn\\
A_3&=&\frac{g}{2\,M}\bigg[\frac{F_2^n}{k\cdot q+p\cdot q}- \frac{F_2^p}{k\cdot q-p\cdot q}\bigg]\,, \ \ A_4= \frac{g}{2\,M}\bigg[\frac{F_2^n}{k\cdot q+p\cdot q}+ \frac{F_2^p}{k\cdot q-p\cdot q}\bigg]\,,\nn\\
A_5 &=& \frac{g}{q^2}\bigg[\frac{F_1^p-F_1^n}{k\cdot q} + \frac{2 F^\pi}{q^2-2\,k\cdot q}\bigg], \ \ A_6=0.
\label{eq:invamp1}
\end{eqnarray}
Denoting  $k$ the four-momentum of the charged meson in the final state and  $p_1\, (p_2),$ the four-momentum of the nucleon (antinucleon), then
the corresponding expressions for the amplitudes $A_i$ in the process $e^+(k_2) +e^-(k_1) \rightarrow \pi^+(k) + n(p_1)+ \bar{p}(p_2)$ can be derived from the last relations by the simple substitution
$F_{1,2}^p \leftrightarrow F_{1,2}^n.$

Note that if
\begin{equation}\label{eq:FF1pi}
 F^{\pi} =  F_1^p - F_1^n, \ J_\mu=\bar{J}_\mu, \ q^\mu\,J_\mu =0, \ A_5 = \frac{g\,(F_1^p-F_1^n)}{k\cdot q(q^2-2\,k\cdot q)},
\end{equation}
the pion formfactor has dipole behaviour at large $q^2$ and this circumstance is against the quark counting rule  \cite{Matveev:1973uz}.

The square of the matrix element can be written as
\begin{equation}\label{eq:M220}
|\mathcal{M}|^2=\frac{16\pi^2\alpha^2}{q^4}L^{\mu\nu}H_{\mu\nu}, \ \
L^{\mu\nu}=l^{\mu}l^{\nu*}, \ \  H_{\mu\nu}=J_{\mu}J_{\nu}^*.
\end{equation}

When the polarizations of the final particles are not measured, the hadron tensor $H_{\mu\nu}$ is:
\begin{equation}\label{eq:HT21}
H_{\mu\nu}(0)=H_1\tilde g_{\mu\nu}+H_2\tilde k_{\mu}\tilde k_{\nu}+H_3\tilde
p_{\mu}\tilde p_{\nu}+H_4(\tilde p_{\mu}\tilde k_{\nu}+\tilde p_{\nu}\tilde
k_{\mu})+iH_5(\tilde p_{\mu}\tilde k_{\nu}-\tilde p_{\nu}\tilde
k_{\mu}),
\end{equation}
where $\tilde g_{\mu\nu}=g_{\mu\nu}-q_{\mu}q_{\nu}/q^2$ and $\tilde
p_{\mu}=p_{1\mu}-(p_\cdot q/q^2)q_{\mu}, \,\, \tilde{k}_\mu = k_\mu - (k\cdot q /q^2) q_\mu $. $H_i$, (i=1-5) are the so-called
structure functions depending on three variables $s=q^2,\,s_1=(p_1+k)^2,\,s_2=(p_2+k)^2$.

The full differential cross section becomes:
\begin{equation}\label{difsec}
d\,\sigma =\frac{\alpha^2}{8\pi^3\,q^6}L^{\mu\nu}\,H_{\mu\nu}\,d\,R_3, \ \ d\,R_3=\frac{\pi}{16(s-2m_e^2)}
\frac{dt_1\,dt_2\,ds_1\,ds_2}{\sqrt{-\Delta}}\,,
\end{equation}
where $\Delta$ is the Gramian determinant \cite{Byckling:1971vca} and $t_1= (k_1-p_1)^2, \, t_2 =(k_2-p_2)^2.$
In our calculation, we can neglect the electron mass with very high accuracy.

When the electron beam is polarized, the leptonic tensor $L_{\mu\nu}$ has the following form
\begin{equation}\label{eq:Lmn}
L_{\mu\nu}=-q^2g_{\mu\nu}+2 (k_{1\mu}k_{2\nu}+k_{1\nu}k_{2\mu})+2im_e (\mu\nu\eta q),
\end{equation}
where $(\mu\nu ab)=\epsilon_{\mu\nu\varrho\sigma}a^{\varrho}b^{\sigma}$ and $\eta_{\mu}$ is the
spin four-vector of the electron (we chose $\epsilon^{0123}=-~\epsilon_{0123}=+1$).

The expression for the hadron tensor in terms of the amplitudes $A_i$ is given by formulas (27$-$31) in Ref. \cite{Gakh:2022fad}.



\section{Calculations of differential distributions}
\subsection{Pion form factor}
The pion electromagnetic form factor $F_\pi(q^2)$ is the subject of both, experimental and theoretical investigations. Experiments of scattering of pions on atomic electrons
$\pi^-+e^-\to \pi^-+e^-$ allowed to determine
the space-like pion form factor (where $-q^2 = Q^2 >0 $) for $Q^2 \le 0.253$, at Fermilab \cite{Dally:1982zk,Dally:1981ur} and CERN \cite{NA7:1986vav}. Larger $Q^2$
values can be reached in the electroproduction cross section $^1 H(e,e'\pi^+)n,$ but with model dependent assumptions. Such measurements were performed at Cornell, DESY \cite{Ackermann:1977rp} and more recently at Jefferson Lab \cite{Horn:2007ug, JeffersonLab:2008jve,JeffersonLab:2008gyl}.

In the timelike region an experiment  CERN \cite{Amendolia:1984nz},  the annihilation reaction $e^+ + e^-\to \pi^+ + \pi^-$
was explored in the $q^2=s=$(0.1- 0.18) GeV$^2$ range.
The cross section for the ISR process $e^+ + e^-\to \pi^+ + \pi^- \,(\gamma)$ by Babar \cite{BaBar:2012bdw} was measured in the $\rho$ resonance region with a precision up to 0.5\%. Other data from the CMD-2 \cite{CMD-2:2006gxt} and from the BESIII collaboration, \cite{BESIII:2015equ} covering a smaller energy range, complete the existing data and overlap in this kinematical region.

Different theoretical approaches and models were developed to describe the pion electromagnetic form factor. Among them one may quote the sum rules of QCD \cite{Nesterenko:1982gc}, perturbative QCD \cite{Bakulev:2004cu,Stefanis:1998dg}, AdS$_5$/CFT correspondence \cite{Brodsky:2006uq,Brodsky:2006uqa}, a model based on saturating Regge trajectories with asymptotical quark counting rules behaviour \cite{Vanderhaeghen:1997ts}. Recently a vector meson dominance fit was extended to pion \cite{Lomon:2016eyp}. Lattice gives also a
description of the electromagnetic structure of pion \cite{Gao:2021xsm}.
The space and timelike behavior of the pion electromagnetic form factor is reproduced in an integral representation based on Green functions in Ref. \cite{Sauli:2022ild}.

\begin{figure}
\centering
\includegraphics[width=0.32\textwidth]{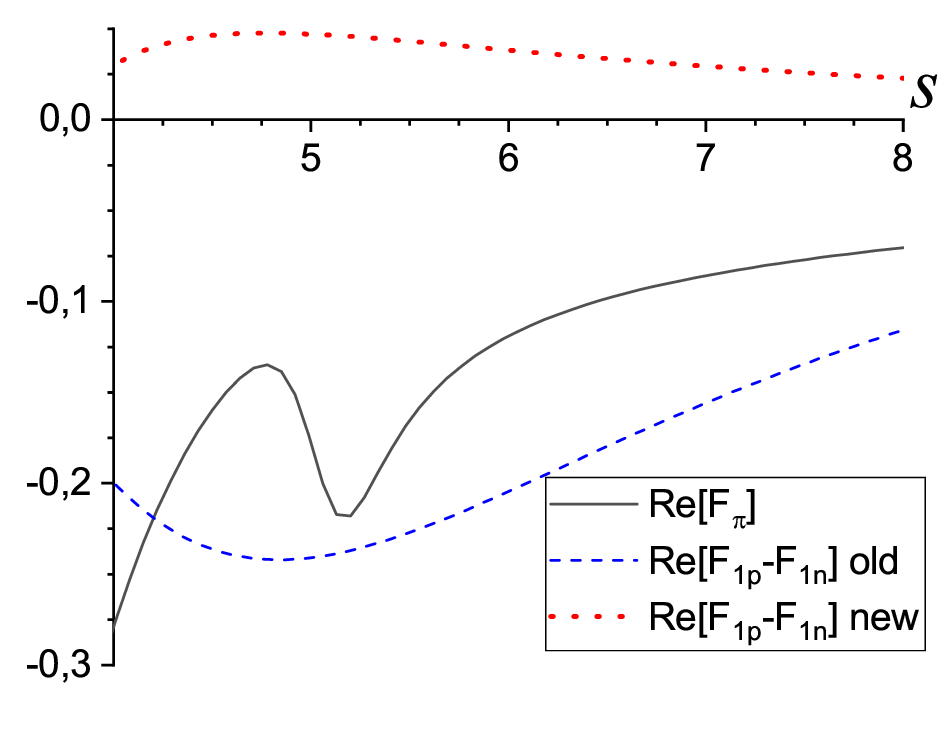}
\includegraphics[width=0.32\textwidth]{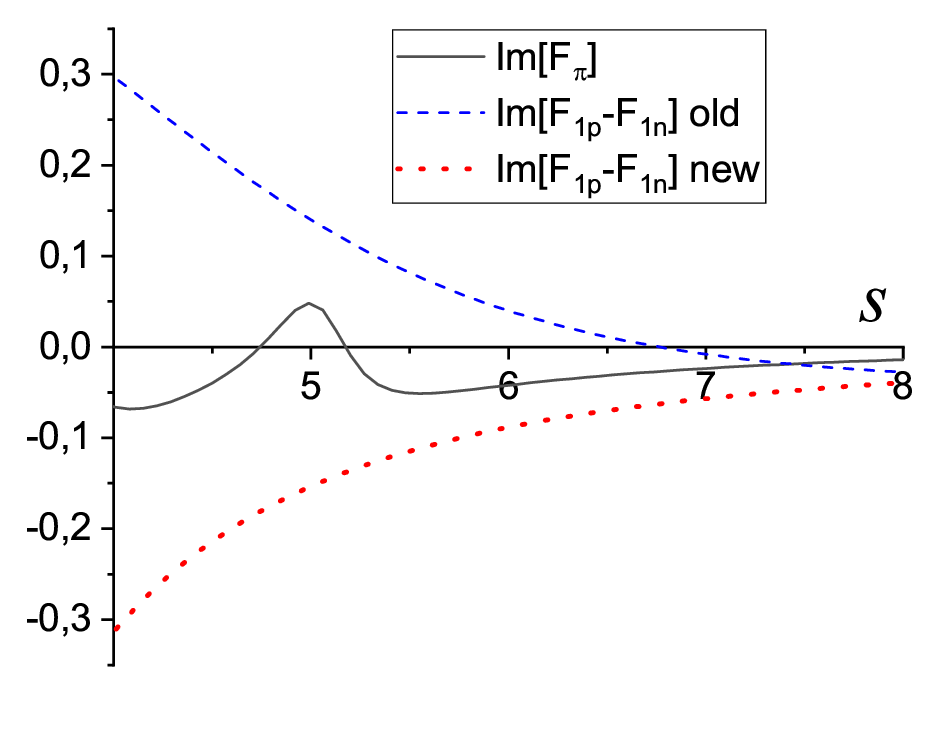}
\includegraphics[width=0.32\textwidth]{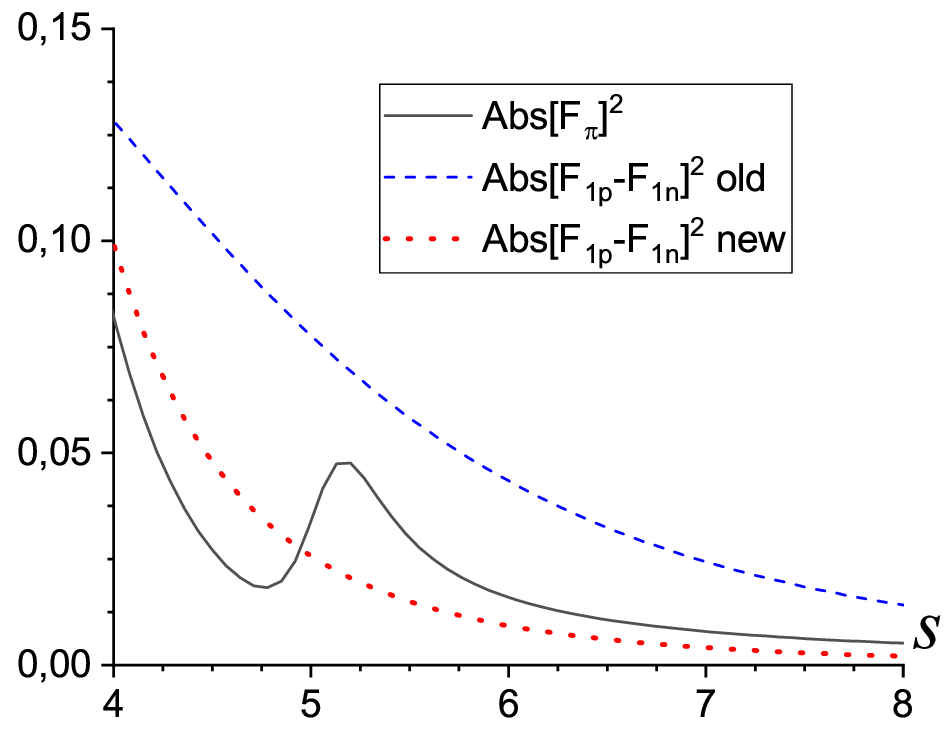}
\vspace{0.3cm}
\includegraphics[width=0.32\textwidth]{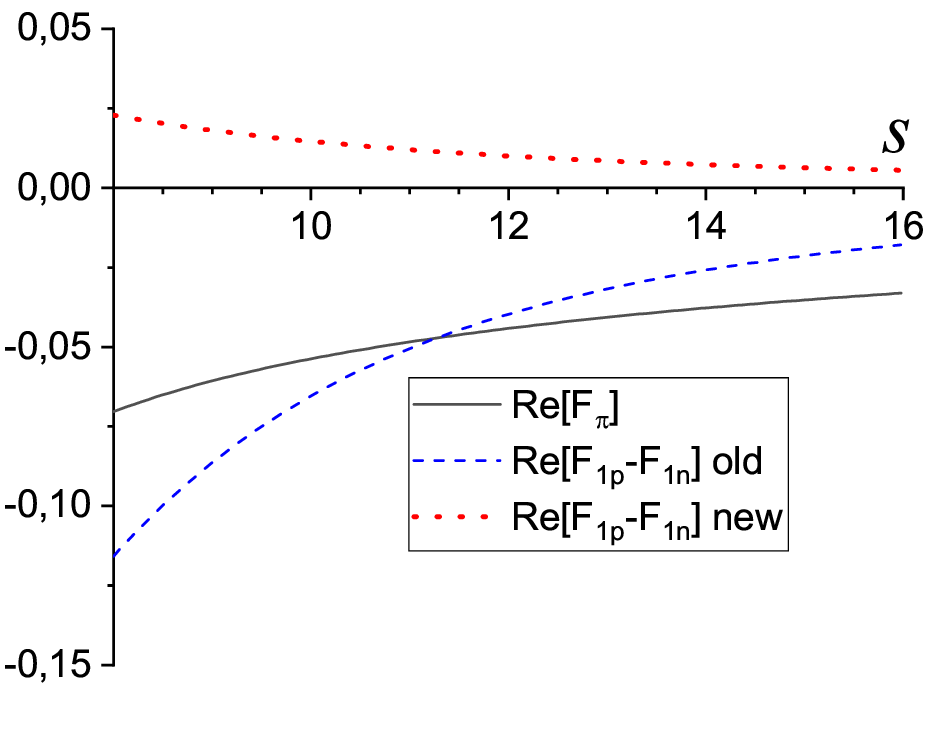}
\includegraphics[width=0.32\textwidth]{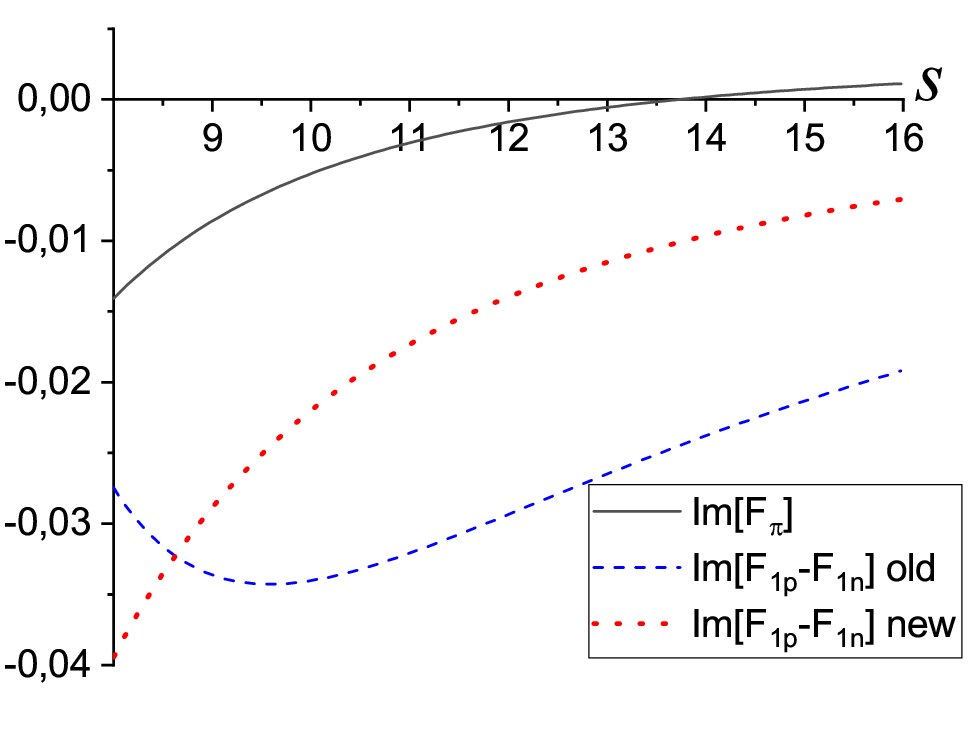}
\includegraphics[width=0.32\textwidth]{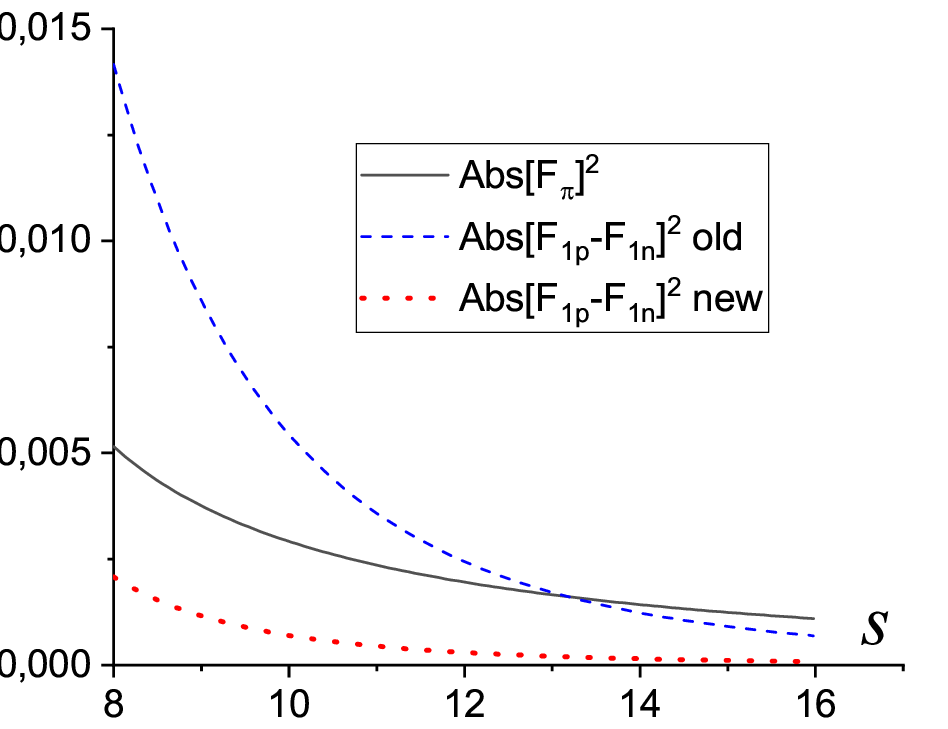}
 \parbox[t]{0.98\textwidth}{\caption{Comparison of the pion formfactor $F_\pi(s)$ defined by the parametrization given in \cite{BaBar:2012bdw}
 (solid curves, monopole behaviour at large $q^2$) with $[F_1^p(s)-F_1^n(s]$ as given by Eq.\,(\ref{eq:FF1pi}). (dipole behaviour at large $q^2.$)}\label{Fig:Fig2}}
\end{figure}

In our numerical calculations we used  a parametrization of the complex pion electromagnetic form factor, based on the VMD with accounting for the contributions of the  family $\rho$, $\rho'$, $\rho''$, $\rho'''$,
as well the $\rho-\omega$ interference, as given in Ref.\cite{BaBar:2012bdw}. The corresponding plots for the real and imaginary parts of the pion form factor as well as its module squared, in the region above
the reaction threshold, are shown in Fig.\ref{Fig:Fig2}. For comparison we also plot the corresponding curves for the pion form factor defined by  Eq.\,(\ref{eq:FF1pi}). In both cases $F_\pi(0) = 1.$ As one can see, the parametrization \cite{BaBar:2012bdw} differs significantly from $[F_1^p(s)-F_1^n(s]$ for both real and imaginary parts as well for the absolute value squared. But the "old" \cite{Iachello:1972nu,Iachello:2004aq} and "new" \cite{Bijker:2004yu} parameterizations for $[F_1^p(s)-F_1^n(s]$ differ even more especially in the region up to $s=8 GeV^2.$

\subsection{Hadron tensor}

For considered here process $e^+\,e^- \rightarrow p\,\bar{n}\,\pi^- $ the hadron structure functions read
\begin{eqnarray}
H_1&=&2 g^2\bigg\{ \big[m^2 s -s_1 s_2 +M^2(s_1+s_2)-M^4\big]\left|\frac{F_1^n}{s_1-M^2}+\frac{F_1^p}{s_2-M^2} \right|^2-\frac{C(s_1,s_2)|F_2^n|^2}{4M^2(s_1-M^2)^2}
\nn\\
&& \ \ - \frac{C(s_2,s_1)|F_2^p|^2}{4M^2(s_2-M^2)^2}-\frac{C_1\,Re[F_2^p\,F_2^{n*}]}{2M^2(s_1-M^2)(s_2-M^2)}
\nn\\&&
 \ \ -C_2(s_1,s_2)\,\bigg[\frac{Re[F_1^n\,F_2^{n*}]}{(s_1-M^2)^2} + \frac{Re[F_1^p\,F_2^{n*}]}{(s_1-M^2)(s_2-M^2)}\bigg]
\nn\\&&
 \ \ -C_2(s_2,s_1)\,\bigg[\frac{Re[F_1^p\,F_2^{p*}]}{(s_2-M^2)^2} + \frac{Re[F_2^p\,F_1^{n*}]}{(s_1-M^2)(s_2-M^2)}\bigg]\bigg\},\nn
\end{eqnarray}
\begin{eqnarray}
C(s_1,s_2 )&=&2M^6 + M^4(s-5s_1-s_2+m^2)-2M^2[m^2(2s+s_1) + s_1(s-2s_1-s_2)]
\nn\\&&
+s_1^2(s-s_1-s_2+m^2),\nn\\
C_1&=&6M^6-M^4[s+5(s_1+s_2)+m^2]+M^2[s_1^2+s_2^2+s(s_1+s_2)+m^2(s_1+s_2-4s) \nn\\&&
+ s_1 s_2(s_1+s_2-s-m^2)],
\nn\\
 \ \ C_2(s_1,\,s_2)&=&2M^4-M^2(3s_1+s_2)-2m^2 s + s_1(s_1+s_2), \nn
\end{eqnarray}
where $C(s_2,\,s_1)$ and $C_2(s_2,\,s_1)$ can be obtained from $C(s_1,\,s_2)$ and $C_2(s_1,\,s_2)$ by the substitution $s_1 \leftrightarrows s_2.$
\begin{eqnarray}
H_2 &=& g^2\Big\{\frac{4}{s_{12}-m^2}\Big[\frac{2 s_{12}|F_\pi|^2}{s_{12}-m^2} + \frac{(2s_{12}-s_1+s_2)\,Re(F_\pi F_{1}^{n*})}{M^2-s_1} \\
&&- \frac{(2s_{12}-s_2+s_1)\,Re(F_\pi F_{1}^{p*})}{M^2-s_2} +\frac{2s\,Re(F_\pi F_{2}^{n*})}{M^2-s_1} - \frac{2s\,Re(F_\pi F_{2}^{p*})}{M^2-s_2}\Big]\nn\\
&&+\frac{2(2M^2+m^2-2s_1)}{(M^2-s_1)^2}\Big(|F_{1}^{n}|^2-\frac{s}{4M^2}|F_{2}^{n}|^2\Big)+\frac{2(2M^2+m^2-2s_2)}{(M^2-s_2)^2}\Big(|F_{1}^{p}|^2-\frac{s}{4M^2}|F_{2}^{p}|^2\Big)\nn\\
&&-\frac{4(s_{12}+s)Re(F_{1}^{p} F_{1}^{n*})}{(M^2-s_1)(M^2-s_2)}-\frac{2s(6M^2-m^2+s_1+s_2)-(s_1-s_2)^2}{2M^2 (M^2-s_1)(M^2-s_2)}Re(F_{2}^{p} F_{2}^{n*})\nn\\
&&-\frac{8s}{(M^2-s_1)(M^2-s_2)}\,Re(F_{1}^{p} F_{2}^{n*}+F_{2}^{p} F_{1}^{n*})\Big\},\nn\\
H_3 &=& g^2\bigg[2m^2\left|\frac{F_1^n}{s_1-M^2}+\frac{F_1^p}{s_2-M^2} \right|^2 -\frac{m^2 s}{2M^2}\bigg(\frac{|F_2^n|^2}{(s_1-M^2)^2}
+ \frac{|F_2^p|^2}{(s_2-M^2)^2}\bigg) \nn\\
 &&+\big[4M^4-4M^2(s_1+s_2) -2m^2 s +(s_1+s_2)^2\big]\frac{Re[F_2^p\,F_2^{n*}]}{2M^2(s_1-M^2)(s_2-M^2)}\bigg],\\
 \label{eq:Had2}
H_4&=&g^2\Big\{-\frac{2(M^2+m^2-s_1)}{(M^2-s_1)^2}\Big(|F_{1}^{n}|^2-\frac{s}{4M^2}|F_{2}^{n}|^2\Big)-\frac{2(M^2+m^2-s_2)}{(M^2-s_2)^2}\Big(|F_{1}^{p}|^2-\frac{s}{4M^2}|F_{2}^{p}|^2\Big)\nn\\
&&
+\frac{2(s_2-s_1)}{(M^2-s_1)(M^2-s_2)}\Big[Re(F_{1}^{p} F_{1}^{n*})-\frac{s_{12}-m^2}{4M^2}Re(F_{2}^{p} F_{2}^{n*})\Big]\nn\\
&&+\frac{2(s_{12}+m^2-s)}{s_{12}-m^2}\Big[\frac{Re(F_\pi F_{1}^{n*})}{M^2-s_1}+\frac{Re(F_\pi F_{1}^{p*})}{M^2-s_2} \Big]\Big\},\nn\\
H_5&=&-\frac{2ig^2(s-s_{12}-m^2)}{(M^2-s_1)(M^2-s_2)}Im\Big(F_{1}^{p} F_{1}^{n*}-\frac{s}{4M^2}F_{2}^{p} F_{2}^{n*}+\frac{s_2-M^2}{s_{12}-m^2}F_\pi F_{1}^{n*}
+\frac{s_1-M^2}{s_{12}-m^2}F_\pi F_{1}^{p*}\Big).\nn
\end{eqnarray}

\subsection{Double differential distribution}

We use approach developed in \cite{Gakh:2022fad} to perform two analytical integrations and obtain the double differential distributions over invariant variables.
By integration over $t_1$ and $t_2$ we can receive $(s_1,\,s_2)$ as well $(s_1,\,s_{12}),\ (s_2,\,s_{12})$-distributions.

The double differential $(s_1,\,s_2)$ distribution are written as follows
\begin{eqnarray}
\frac{d\,\sigma}{ds_1\,ds_2}&=&\frac{g^2\,\alpha^2}{8\pi\,s^3}\,W_2(s_1,s_2),\label{eq:Dss1s2-22}\\
W_2(s_1,s_2) &= &\frac{1}{3s}\big[D_{1n}|F_1^n|^2+D_{1p}|F_1^p|^2 + D_\pi|F_\pi|^2 \label{eq:2W2s1s2} \\
&&
+ 2Re\big(D_{11}F_1^p F_1^{n*}+D_{1p}^\pi F_\pi F_1^{p*}-
D_{1n}^\pi F_\pi F_1^{n*}\big)\big] \nn \\
&&+\frac{1}{3}Re\Big(D_{21}^{pn}F_2^p F_1^{n*} + D_{12}^{pn}F_1^p F_2^{n*}\Big) + Re\big(D_{12}^n F_1^n F_2^{n*} + D_{12}^p F_1^p F_2^{p*}\big)\nn\\
&&+\frac{1}{12 M^2}\big[D_{2n}|F_2^n|^2+D_{2p}|F_2^p|^2+2 D_{22} Re(F_2^p F_2^{n*})\big]\nn\\
&&+\frac{(s_{12}-s-m^2)^2-4 m^2s}{3(s_{12}-m^2)}Re\Big(\frac{F_\pi F_2^{p*}}{s_2-M^2}-\frac{F_\pi F_2^{n*}}{s_1-M^2}\Big),\nn
\end{eqnarray}
where the real coefficients are
\begin{eqnarray}
D_{1p} &=& s_{12} + \frac{2 s}{s_2-M^2}\Big[s_1-M^2-\frac{m^2(2M^2+s)}{s_2-M^2}\Big], \ \ D_{11} = 2s -s_{12} + \frac{2 m^2s(s_{12}-s-2M^2)}{(s_1-M^2)(s_2-M^2)}, \nn\\
D_{22}&=& 4M^2-s -\frac{2(M^4-M^2s_2-m^2s)}{s_1-M^2}-\frac{2(M^4-M^2s_1-m^2s)}{s_2-M^2}-\frac{m^2s(2m^2+s)}{(s_1-M^2)(s_2-M^2)}, \nn\\
D_{21}^{pn} &=& \frac{s_2-s_1}{s_2-M^2}+\frac{2(s_2-M^2)}{s_1-M^2} - \frac{2 m^2s}{(s_1-M^2)(s_2-M^2)}, \nn\\
D_{2p} &=& 2s_{12} + \frac{s(s_1-M^2)}{(s_2-M^2)} - \frac{m^2s(8M^2+s)}{(s_2-M^2)^2}, \ \ D_{12}^p = 1 + \frac{s_1-M^2}{s_2-M^2} - \frac{2 m^2s}{(s_2-M^2)^2},\nn\\
D_\pi&=& \frac{s_{12}}{(s_{12}-m^2)^2}\big[(s_{12}-s-m^2)^2-4 m^2s\big], \ \ D_{1p}^\pi = -s_{12} + \frac{s s_{12}}{(s_{12}-m^2)}\Big(1-\frac{2 m^2}{s_2-M^2}\Big).\nn
\end{eqnarray}
The coefficients $D_{1n}$, $ D_{12}^{pn}$, $ D_{12}^{n}$,  $D_{2n}$ and $D_{1n}^\pi$ can be obtained from $D_{1p}$, $D_{21}^{pn}$, $ D_{12}^{p}$,  $D_{2p}$ and $D_{1p}^\pi$ with the change $s_1\rightleftarrows s_2$.

The distributions over $(s_1,\,s_{12})$  and $(s_2,\,s_{12})$ can be obtained from (\ref{eq:Dss1s2-22}) by substitution $s_2=2M^2+m^2+s-s_1-s_{12}$ at fixed
 $s_1$ and $s_1=2M^2+m^2+s-s_2-s_{12}$ at fixed $s_2,$ respectively. The analytic form of the double differential distributions over $(s_1,\,t_2)$ and $(s_2,\,t_1)$  are very cumbersome and we do not give them in this article. The $(t_1,\,t_2)$ distribution is calculated by numerical integration.

\begin{figure}[h!]
\centering

\includegraphics[width=0.26322\textwidth]{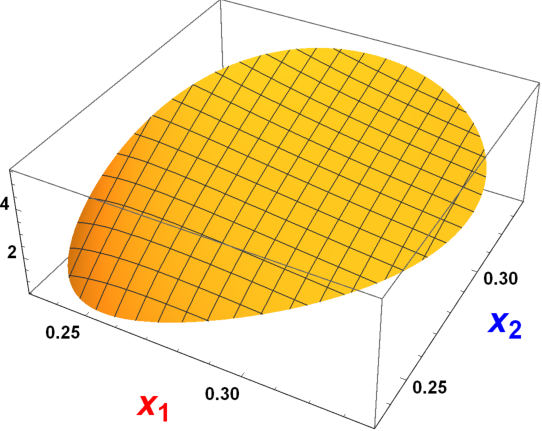}
\hspace{0.5cm}
\includegraphics[width=0.26322\textwidth]{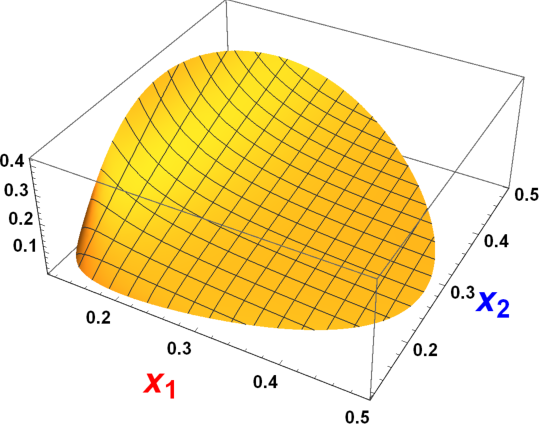}
\hspace{0.5cm}
\includegraphics[width=0.26322\textwidth]{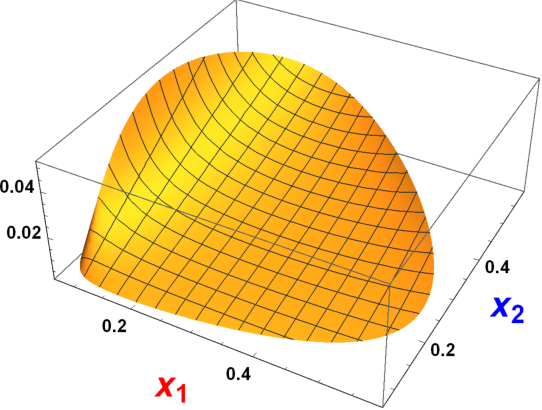}
\vspace{0.15cm}
\includegraphics[width=0.25\textwidth]{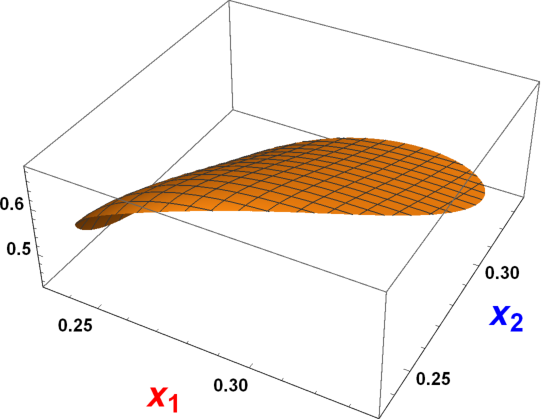}
\hspace{0.5cm}
\includegraphics[width=0.25\textwidth]{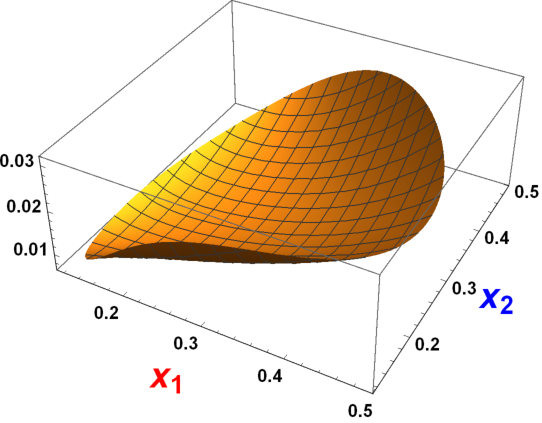}
\hspace{0.5cm}
\includegraphics[width=0.25\textwidth]{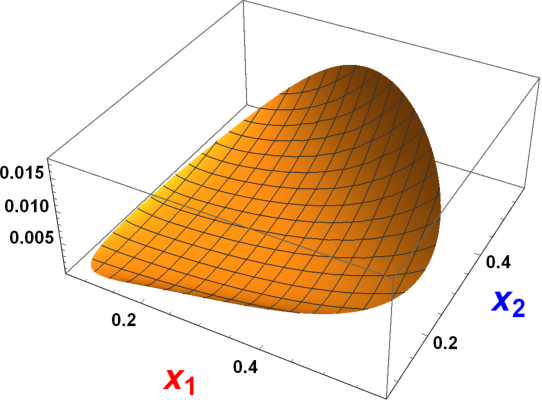}
\vspace{0.15cm}
\includegraphics[width=0.25\textwidth]{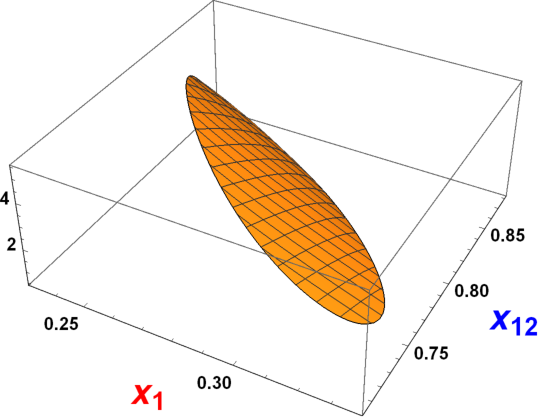}
\hspace{0.5cm}
\includegraphics[width=0.25\textwidth]{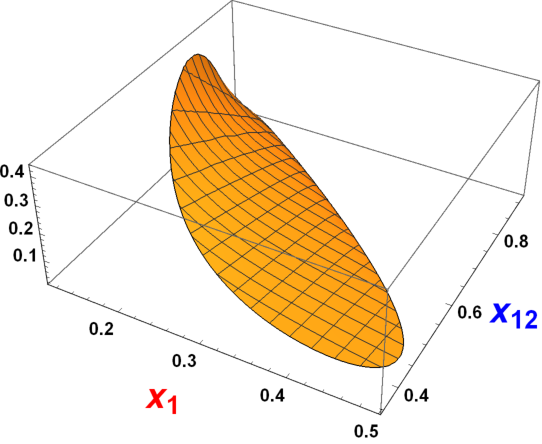}
\hspace{0.5cm}
\includegraphics[width=0.25\textwidth]{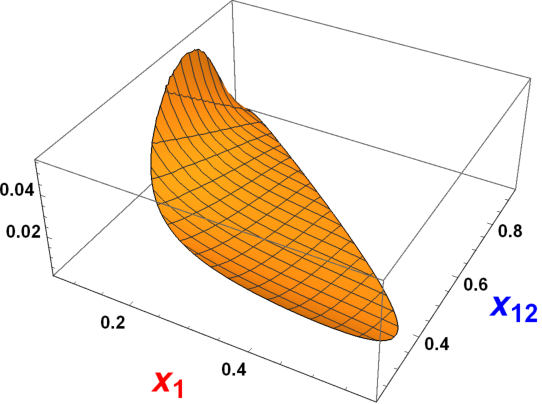}
\vspace{0.15cm}
\includegraphics[width=0.25\textwidth]{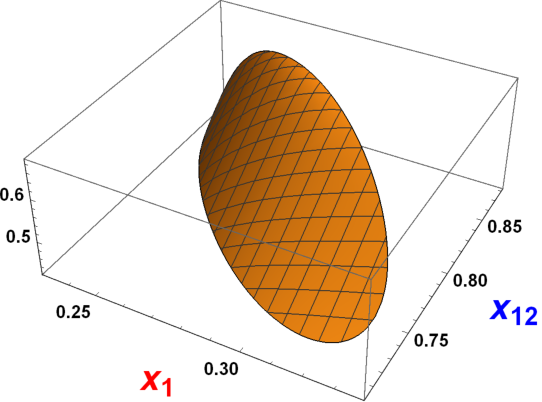}
\hspace{0.5cm}
\includegraphics[width=0.25\textwidth]{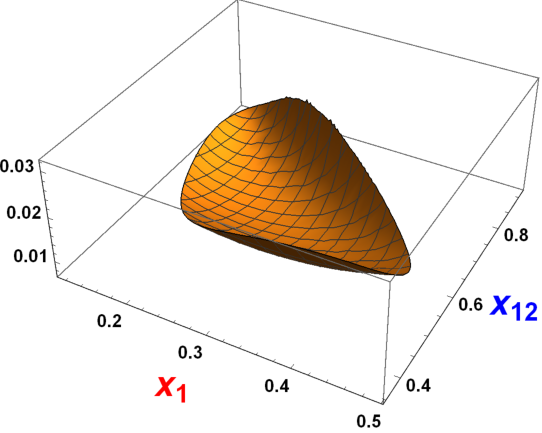}
\hspace{0.5cm}
\includegraphics[width=0.25\textwidth]{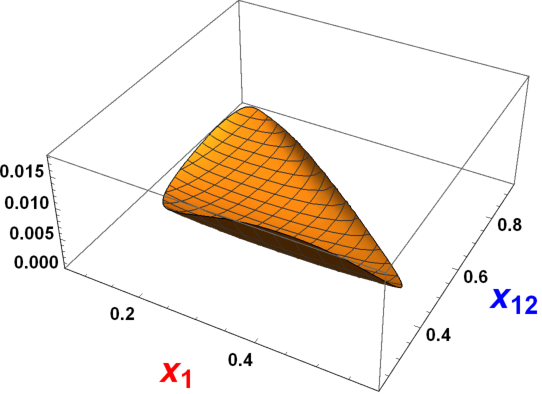}
\vspace{0.15cm}
\includegraphics[width=0.25\textwidth]{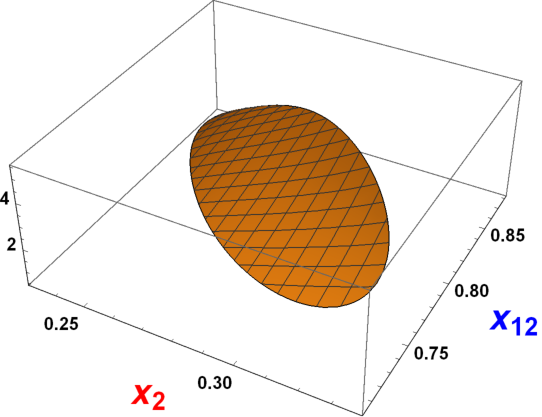}
\hspace{0.5cm}
\includegraphics[width=0.25\textwidth]{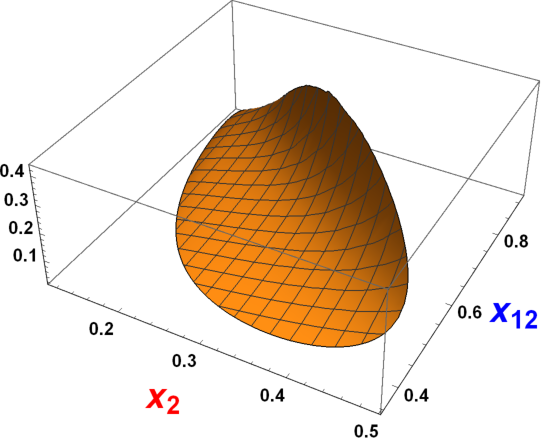}
\hspace{0.5cm}
\includegraphics[width=0.25\textwidth]{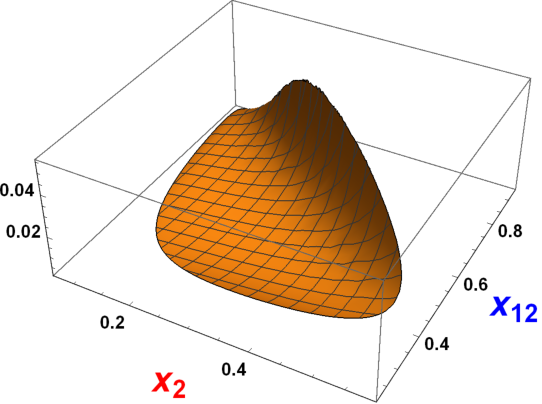}
\vspace{0.15cm}
\includegraphics[width=0.25\textwidth]{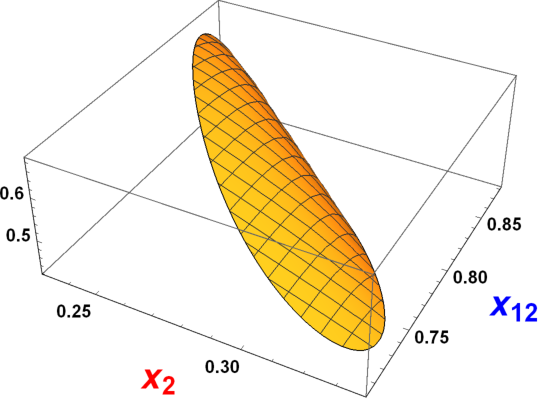}
\hspace{0.5cm}
\includegraphics[width=0.25\textwidth]{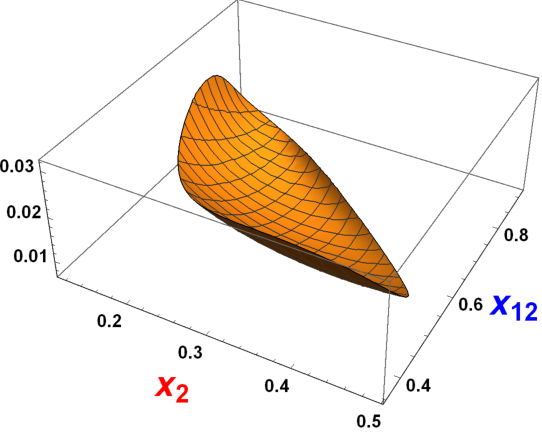}
\hspace{0.5cm}
\includegraphics[width=0.25\textwidth]{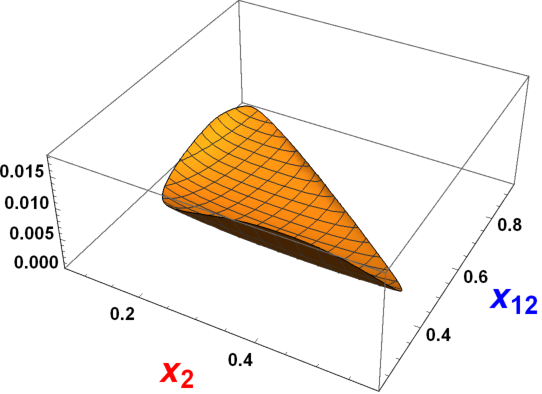}
\vspace{0.15cm}
{ \caption{Double differential distributions (in {\it nb}) over the dimensionless  variables $x_{1,2}=s_{1,2}/s$, $x_{12}=s_{12}/s $ in the reaction  $e^+ + e^- \to p + \bar{n} +\pi^-$ at $s$= 5, 10, 16 GeV$^2 $ (from left to right). The odd (even) rows correspond to the "old" ("new") parametrization of the nucleon form factors.}
\label{Fig:Fig3} }
\end{figure}

\begin{figure}[h!]
\centering

\includegraphics[width=0.26322\textwidth]{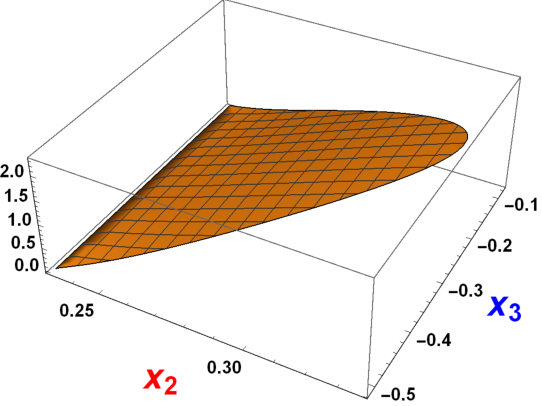}
\hspace{0.5cm}
\includegraphics[width=0.26322\textwidth]{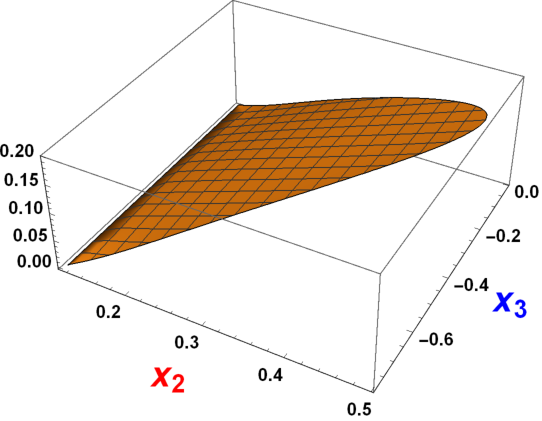}
\hspace{0.5cm}
\includegraphics[width=0.26322\textwidth]{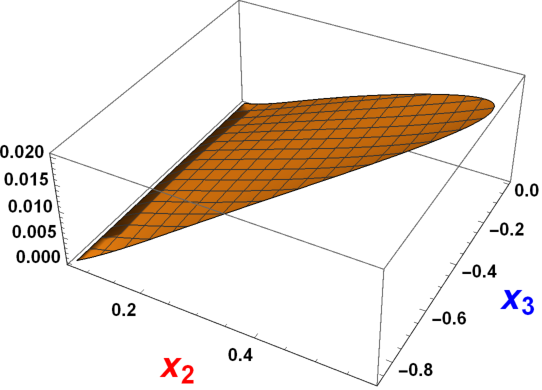}
\vspace{0.15cm}
\includegraphics[width=0.26322\textwidth]{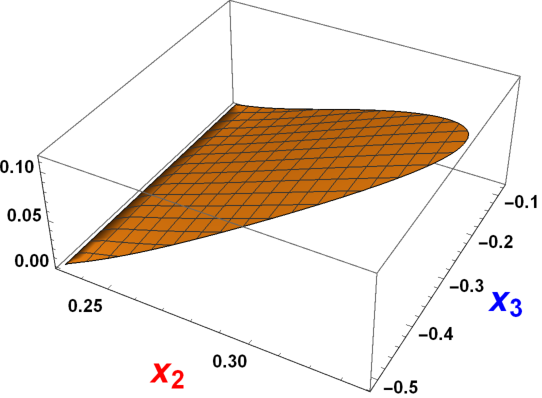}
\hspace{0.5cm}
\includegraphics[width=0.26322\textwidth]{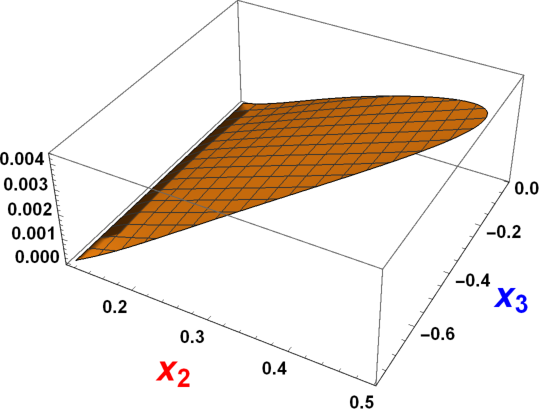}
\hspace{0.5cm}
\includegraphics[width=0.26322\textwidth]{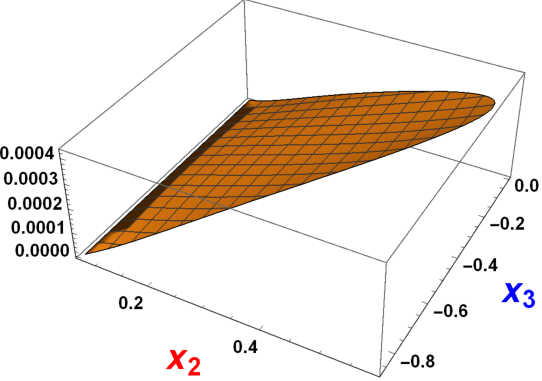}
\vspace{0.15cm}
\includegraphics[width=0.26322\textwidth]{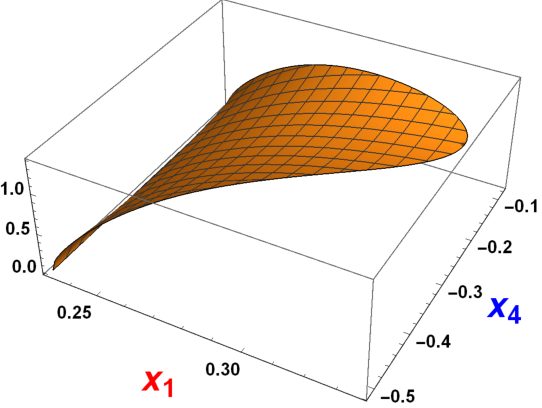}
\hspace{0.5cm}
\includegraphics[width=0.26322\textwidth]{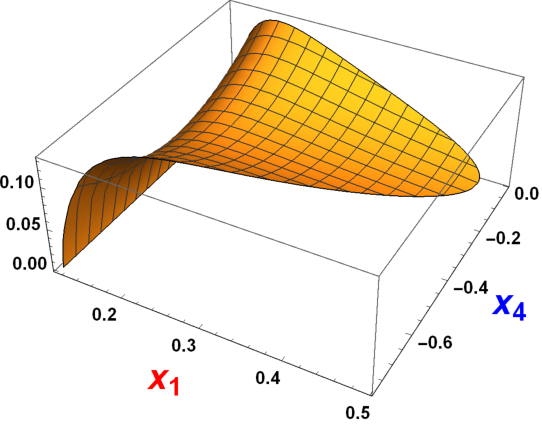}
\hspace{0.5cm}
\includegraphics[width=0.26322\textwidth]{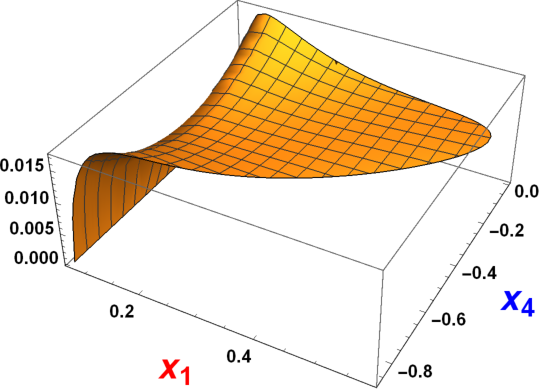}
\vspace{0.15cm}
\includegraphics[width=0.26322\textwidth]{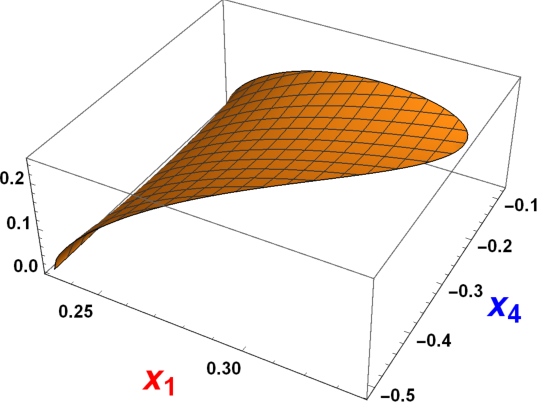}
\hspace{0.5cm}
\includegraphics[width=0.26322\textwidth]{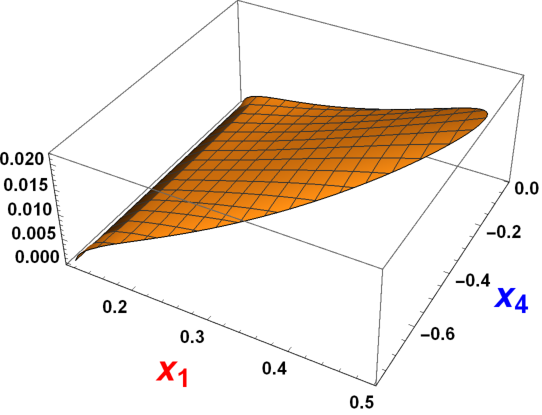}
\hspace{0.5cm}
\includegraphics[width=0.26322\textwidth]{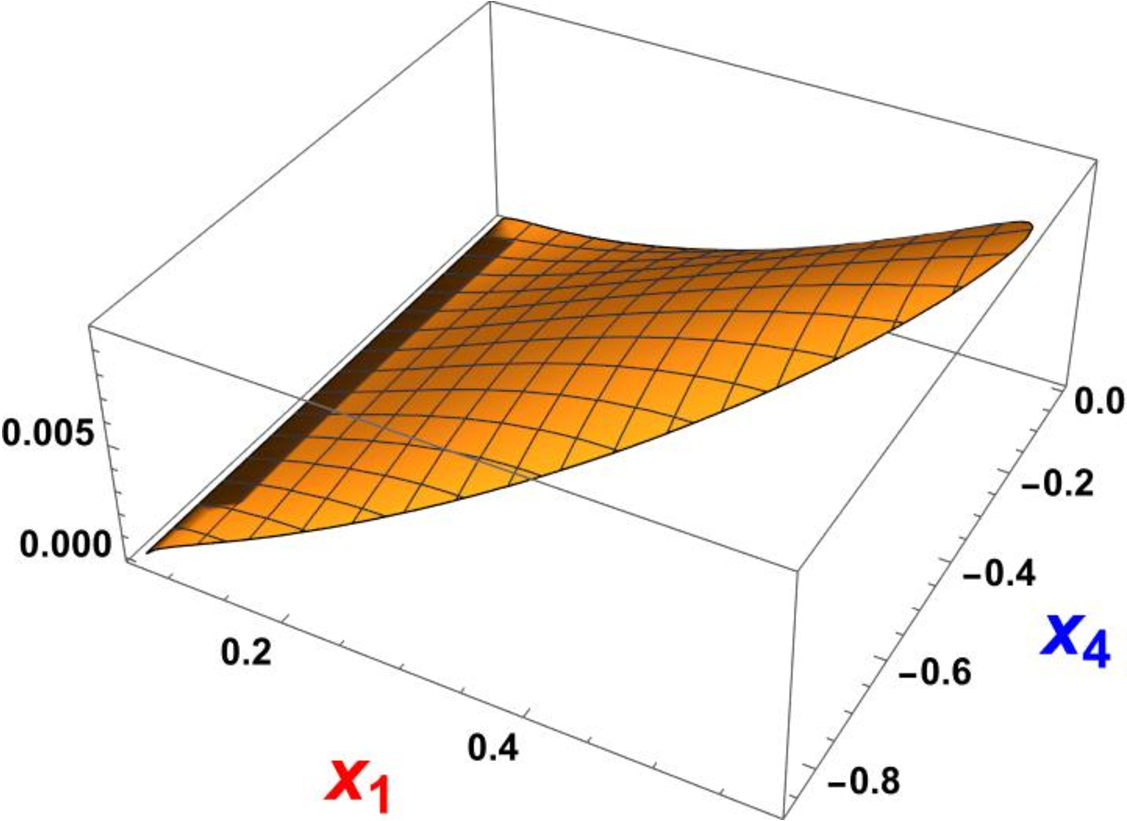}
\vspace{0.15cm}
\includegraphics[width=0.25\textwidth]{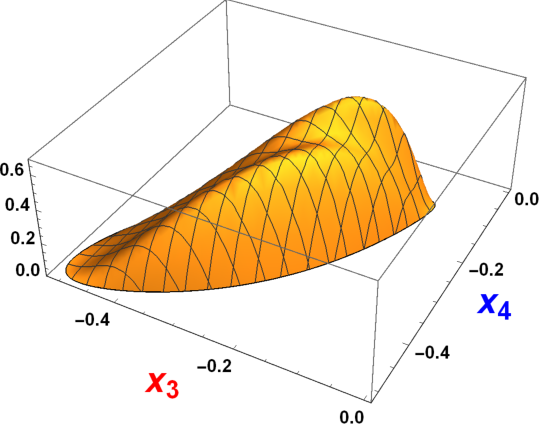}
\hspace{0.5cm}
\includegraphics[width=0.25\textwidth]{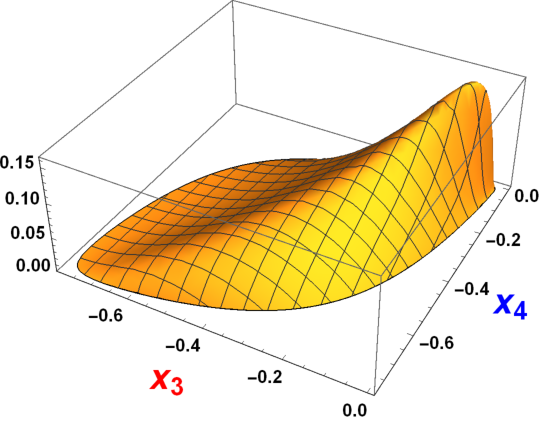}
\hspace{0.5cm}
\includegraphics[width=0.25\textwidth]{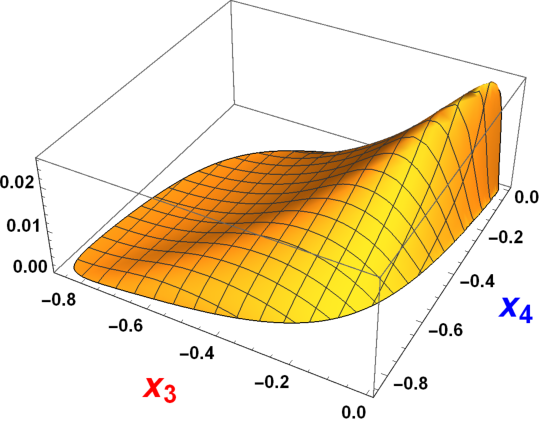}
\vspace{0.15cm}
\includegraphics[width=0.25\textwidth]{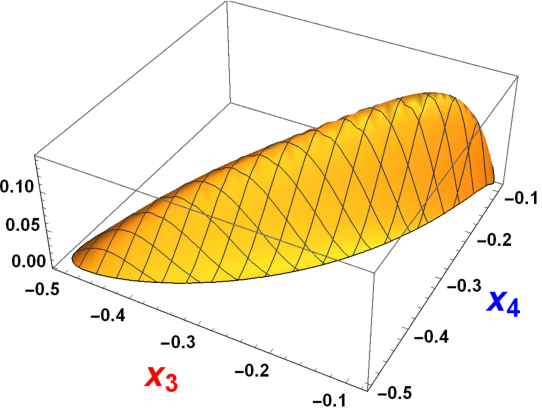}
\hspace{0.5cm}
\includegraphics[width=0.25\textwidth]{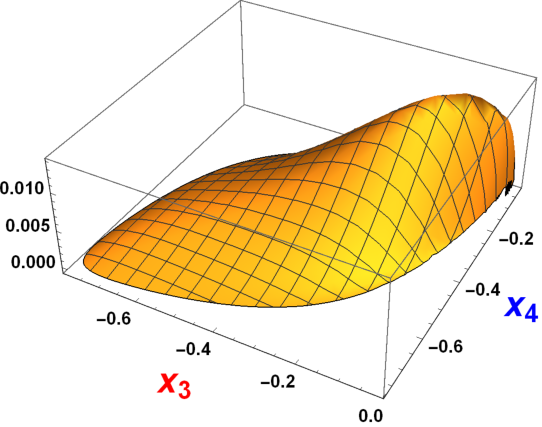}
\hspace{0.5cm}
\includegraphics[width=0.25\textwidth]{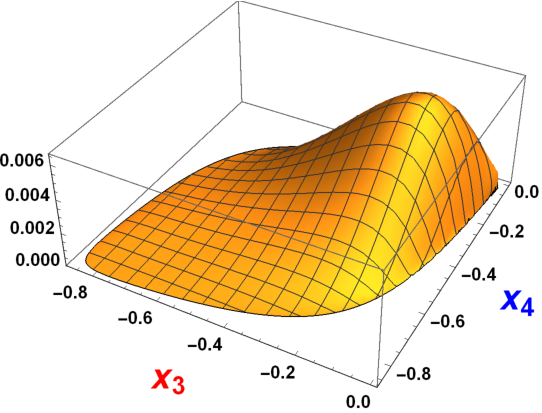}
\vspace{0.15cm}
{ \caption{The same as in Fig.\ref{Fig:Fig3} but for double differential distributions with the dimensionless  variables $x_{3,4}=t_{3,4}/s$. }
\label{Fig:Fig4} }
\end{figure}

As we noted above, the transition to the reaction $e^+ + e^- \to n + \bar{p} +\pi^+$ can be done by substitution of the
form factors $F_{1,2}^p,\rightleftarrows F_{1,2}^n,\,F_\pi \to - F_\pi$. In the analytical calculations we showed that this substitution (on the level of the full differential cross section) is equivalent to change $s_1\rightleftarrows s_2,\, t_1\rightleftarrows t_2$. Moreover, the phase space $d R_3$ and the limits of variations of the invariant variables are symmetrical under this change.
Therefore we conclude that after integration over the variables $t_1$ and $t_2$ the $(s_1,\,s_2)$ distribution in the process $e^+ + e^- \to n + \bar{p} +\pi^+$ coincides with the $(s_2,\,s_1)$ distribution in the process $e^+ + e^- \to p + \bar{n} +\pi^-.$ It means that the total cross sections of these two reactions are the same. For other double differential distributions, the corresponding transitions to the reaction $e^+ + e^- \to n + \bar{p} +\pi^+$ are
\begin{equation}\label{eq:substitutions}
(t_1,t_2)\to (t_2,t_1),\,(s_1,t_2)\to (s_2,t_1),\,(s_2,t_1)\to (s_1,t_2),
\end{equation}
\[(s_1,s_{12})\to (s_2,s_{12}),\,(s_2,s_{12})\to (s_1,s_{12}).\]
The double differential distributions over different pairs of dimensionless invariant variables are shown in Figs.\ref{Fig:Fig3}, \ref{Fig:Fig4}.

\subsection{Single differential distributions}

Integrating differential cross section (\ref{eq:Dss1s2-22}) over the variable $s_2$ in limits $s_{2-}\leq s_2 \leq s_{2+},$ where \cite{Gakh:2022fad}
\begin{equation}\label{eq:s2pm}
s_{2\pm} = \frac{1}{2s_1}\big[D(s,s_1) \pm \sqrt{F(s,s_1)\,G(s,s_1)}\big],
\end{equation}
\[D(s,s_1)=M^4-M^2(s-2s_1+m^2) + m^2(s+s_1)+s_1(s-s_1),\]
\[F(s,s_1)=(s+s_1-M^2)^2 - 4s\,s_1, \ G(s,s_1)= (s_1+m^2-M^2)^2-4m^2s_1,\]
the following distribution over $s_1$ is obtained
\begin{equation}
\frac{d\,\sigma}{d s_1} = \frac{g^2\,\alpha^2}{8\,\pi\,s^3}\,{W}_1(s_1),\label{eq:2Dss1}
\end{equation}
\[{W}_1(s_1)=K\Big(\frac{|F_1^n|^2}{6 s}{D}_{1n} + \frac{|F_2^n|^2}{12 M^2}{D}_{2n}\Big) + \frac{|F_1^p|^2}{6s}\big[K\,{D}_{1p}
-4s(s_1-M^2)L_N\big]\]
\[+ \frac{|F_2^p|^2}{12M^2}\big[K\,{D}_{2p}-s(s_1-M^2)\,L_N\big]\]
\[+ \frac{Re(F_1^p\,F_1^{n*})}{3 s}\Big[-\frac{K\,{D}_{11}}{s_1^2(s_1-M^2)}+\frac{8m^2 s(s_1+M^2-m^2)}{s_1-M^2}L_N\Big]\]
\[+\frac{Re(F_2^p\,F_1^{n*})}{3 (s_1-M^2)}\Big[\frac{K}{s_1^2}{D}_{21}^{pn} +\big[(s_1-M^2)^2+2m^2 s\big]L_N\Big]
+\frac{Re(F_1^n\,F_2^{n*})}{2s_1^2(s_1-M^2)^2} K{D}_{12}^{n}\]
\[+\frac{Re(F_1^p\,F_2^{n*})}{6 (s_1-M^2)}\Big[\frac{K}{s_1^2}{D}_{12}^{pn} +4\big[m^2 s-(s_1-M^2)^2\big]L_N\Big]
 +Re(F_1^p\,F_2^{p*}) \big[K\,{D}_{12}^{p}  - (s_1-M^2)L_N \big]\]
\[+\frac{Re(F_2^p\,F_2^{n*})}{6 M^2}\Big[\frac{K}{s_1^2}{D}_{22} +\Big(2M^4-2M^2 s_1 - 2m^2 s +\frac{m^2 s(s+2m^2)}{s_1-M^2}\Big)L_N \Big]\]
\[+\frac{|F_\pi|^2}{6s}\big[K\,{D}^\pi +2s(s-6m^2)L_\pi\big] +\frac{Re(F_\pi\,F_1^{n*})}{3s}\Big[K\,{D}_{1n}^\pi +
\frac{2m^2 s(M^2+2m^2-s_1)}{s_1-M^2}L_\pi\Big] \]
\[+\frac{Re(F_\pi\,F_1^{p*})}{3s}\Big[\frac{K\,{D}_{1p}^\pi}{s_1^2} + \frac{2m^2 s}{s-s_1+M^2}\big[ 2(s-s_1+M^2+m^2)L_N \]
\[+(s-s_1+M^2-2m^2)L_\pi\big]\Big]+\frac{Re(F_\pi\,F_2^{n*})}{6(s_1-M^2)}\Big[\frac{K}{s_1^2}{D}_{2n}^\pi -2s(s-4 m^2)L_\pi\Big]\]
\begin{equation}\label{eq:2W1s1}
+\frac{Re(F_\pi\,F_2^{p*})}{3(s-s_1+M^2)}\Big[-\frac{K (s-s_1+M^2)}{s_1} +\big[4m^2s-(s_1-M^2)^2\big]L_N + s(s-4m^2)L_\pi\Big],
\end{equation}
where
\[K= \sqrt{[M^4-2M^2(s_1+m^2)+(s_1-m^2)^2][M^4-2M^2(s_1+s)+(s_1+s)^2]},\]
\[L_N=\ln\frac{l_n^2}{4 s_1 d_N}, \ L_\pi =\ln\frac{l_\pi^2}{4s_1 d_\pi},\]
\[d_N=M^6 -2M^4s_1 +M^2(s_1^2-3m^2s)+m^2s(s-s_1+m^2),\]
\[d_\pi=M^2s^2+m^2[M^4-M^2(3s+2s_1)+m^2s-s_1(s-s_1)],\]
\[l_N=M^4-M^2(s+m^2)+m^2(s+s_1)+s_1(s-s_1) -K, \ l_\pi= l_N-2s_1(s-s_1+M^2),\]
\[{D}_{1n}=\frac{2m^2s(4M^2+3s)}{M^4(s_1-M^2)}-\frac{4m^2s(2M^2+s)}{M^2(s_1-M^2)^2}+\frac{2M^6-M^4(s-m^2)-m^2s(8M^2+6s)}{M^4 s_1}\]
\[+\frac{(M^2-m^2)(2s^2-M^2 s-M^4)}{M^2s_1^2} -1,\]
\[{D}_{2n}=\frac{m^2s(16 M^2+3s)}{2M^4(s_1-M^2)}-\frac{m^2s(8M^2+s)}{M^2(s_1-M^2)^2}+\frac{4M^6+M^4(s+2m^2)-m^2s(16 M^2+3s)}{2M^4 s_1}\]
\[+\frac{(M^2-m^2)(s^2+M^2 s-2M^4)}{2M^2s_1^2} -1,\]
\[{D}_{1p}=\frac{-4m^2 s(s+2M^2)}{d_N} +\frac{(M^2-m^2)(s-M^2)}{s_1^2} +\frac{s+2M^2+m^2}{s_1}-1,\]
\[{D}_{2p}=\overline{D}_{1p} +\frac{3m^2 s^2}{d_N},\]
\[{D}_{11} = M^6 -M^4(s+3s_1+m^2)+ M^2\big[m^2s + s_1(4s+3s_1)\big] -s_1(s_1-m^2)(3 s+s_1),\]
\[{D}_{21}^{pn}=M^4-M^2(s+s_1+m^2)+m^2(s+s_1)+s s_1,\]
\[{D}_{12}^{n}=-M^6 +M^4(s +3s_1+m^2) -M^2\big[m^2(s+2s_1)+s_1(2s+3s_1)\big]+s_1\big[m^2(s_1-3s)+s_1(s+s_1)\big],\]
\[{D}_{12}^{pn}=-M^4 +M^2(s-2s_1+m^2) - s_1(s-3s_1) - m^2(s+s_1),\]
\[{D}_{12}^{p}= -\frac{1}{s_1} +\frac{2m^2s}{d_N},\]
\[{D}_{22} = -M^4+M^2(s+3s_1+m^2)- s(s_1-2m^2)+\frac{3 M^2 m^2s}{s_1-M^2},\]
\[{D}^\pi = \frac{2m^2s(s-4m^2)}{d_{\pi}}+\frac{(M^2-m^2)(s-M^2)}{s_1^2}+\frac{2M^2+m^2-3s}{s_1} -1,\]
\[{D}^\pi_{1n} = \frac{(M^2-m^2)(s-M^2)}{s_1^2}+\frac{2M^4-M^2(s-m^2)-4m^2s}{M^2 s_1} + \frac{4m^2s}{M^2(s_1-M^2)} -1,\]
\[{D}^\pi_{1p}= (s_1-M^2)^2 +s(s_1-M^2) + m^2(s-s_1-M^2),\]
\[{D}^\pi_{2n} = (s_1-M^2)^2 +s(3s_1-M^2) + m^2(s+s_1-M^2).\]

The analytical expression of $d\sigma/ds_2$ can be derived from (\ref{eq:2Dss1}) by the changes  $F^p_{1,2} \rightleftarrows F^n_{1,2},\, F_\pi \to - F_\pi, \ s_1\to s_2.$

The distribution over the variable $s_{12}$ is symmetrical under substitution $F^p_{1,2} \rightleftarrows F^n_{1,2},\, F_\pi \to - F_\pi.$ To obtain it one has to integrate the $(s_1,\,s_{12})$ distribution over $s_1$ in the limits $s_{1-}\leq s_1 \leq s_{1+},$ where \cite{Gakh:2022fad}
\begin{equation}\label{eq:s1pm}
s_{1\pm}=\frac{1}{2}\bigg[2M^2+m^2+s-s_1-s_{12} \pm \sqrt{\Big(1-\frac{4M^2}{s_{12}}\Big)\big[(s+m^2-s_{12})^2-4m^2 s\big]}\bigg].
\end{equation}
The result reads
The distribution over the variable $s_{12}$ is symmetrical under substitution $F^p_{1,2} \rightleftarrows F^n_{1,2},\, F_\pi \to - F_\pi$ and has the following form
\begin{equation}\label{eq:2Dss12}
\frac{d\,\sigma}{d s_{12}} = \frac{g^2\,\alpha^2}{8\,\pi\,s^3}\,{W}_1(s_{12}),
\end{equation}
\[{W}_1(s_{12})=\frac{|F_1^n|^2 +|F_1^p|^2}{3s}\Big[\frac{\overline{K}\,{G}_1}{s_{12}\,Y} + 2s(s-s_{12}+m^2)L_{12}\Big]\]
\[+\frac{|F_2^n|^2 +|F_2^p|^2}{12 M^2}\Big[\frac{-\overline{K}\,{G}_2}{s_{12}\,Y} + s(s-s_{12}+m^2)L_{12}\Big]
+\frac{\overline{K}^3|F_{\pi}|^2}{3ss_{12}(s_{12}-4M^2)(s_{12}-m^2)^2}\]
\[+\frac{2 Re(F_1^p \,F_1^{n*})}{3s}\Big[\frac{(2s-s_{12})\overline{K}}{s_{12}}-\frac{4m^2 s (s-s_{12}+2M^2)}{s-s_{12}+m^2}L_{12}\Big]\]
\[+\frac{Re(F_2^p \,F_2^{n*})}{6M^2}\Big[-\frac{s}{s_{12}}\overline{K} +\frac{2\big[2M^2(s-s_{12}+m^2)^2+m^2s(s-2s_{12})\big]}{s-s_{12}+m^2}L_{12} \Big]\]
\[+Re\big[F_1^n\,F_2^{n*} + F_1^p\,F_2^{p*}\big]\Big[-\frac{2 m^2 s \overline{K}}{Y} + (s-s_{12}+m^2)L_{12}\Big]\]
\[+ Re\Big[\frac{F_1^p\,F_2^{n*}+F_2^p\,F_1^{n*}}{3(s-s_{12}+m^2)} +\frac{F_\pi(F_2^{p*}-F_2^{n*})}{3(s_{12}-m^2)}\Big]\big[m^4 +(s-s_{12})^2-2m^2(s+s_{12})\big]L_{12}\]
\begin{equation}\label{eq:2Ws12}
+\frac{2ReF_\pi(F_1^{n*}-F_1^{p*})}{3 s (s_{12}-m^2)}\big[(s_{12}-s-m^2) \overline{K} + 2m^2 s s_{12} L_{12}\big],
\end{equation}
where
\[\overline{K}=\sqrt{s_{12}(s_{12}-4M^2)\big[(s-s_{12})^2-2m^2(s+s_{12})+m^4\big]},\]
\[L_{12} = \ln\frac{[s_{12}(s-s_{12}+m^2)+\overline{K}]^2}{4s_{12} d_{12}}, \ d_{12} = M^2\big[(s-s_{12})^2-2m^2(s+s_{12})+m^4\big]+m^2 s s_{12},\]
\[Y=M^2\big[m^4 +(s-s_{12})^2 -2m^2(s+s_{12})\big] + m^2 s s_{12},\]
\[{G}_1 = M^2\big[m^4(s_{12}-2s)+m^2(4s^2-2s  s_{12} -2 s_{12}^2) -(2s-s_{12})(s-s_{12})^2\big] +m^2s s_{12}(s_{12} - 4s),\]
\[{G}_2 = M^2\big[m^4(s-2s_{12})-2m^2(s^2-5s  s_{12} -2 s_{12}^2) +(s-2s_{12})(s-s_{12})^2\big] +2m^2s s_{12}(s - s_{12}).\]

\begin{figure}[h!]
\centering
\includegraphics[width=0.262\textwidth]{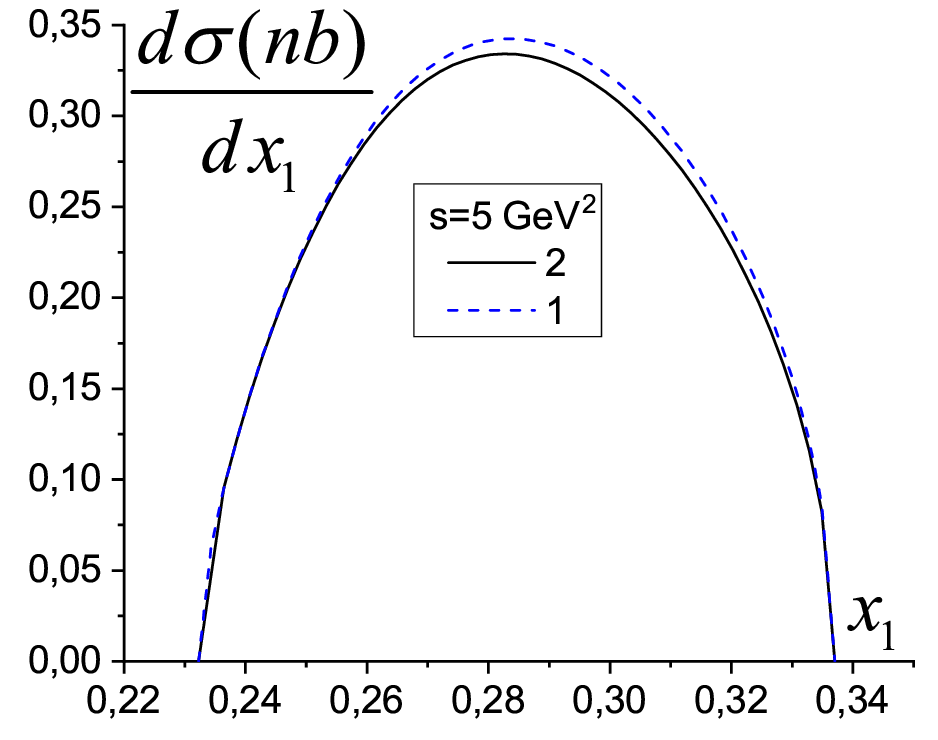}
\hspace{0.15cm}
\includegraphics[width=0.262\textwidth]{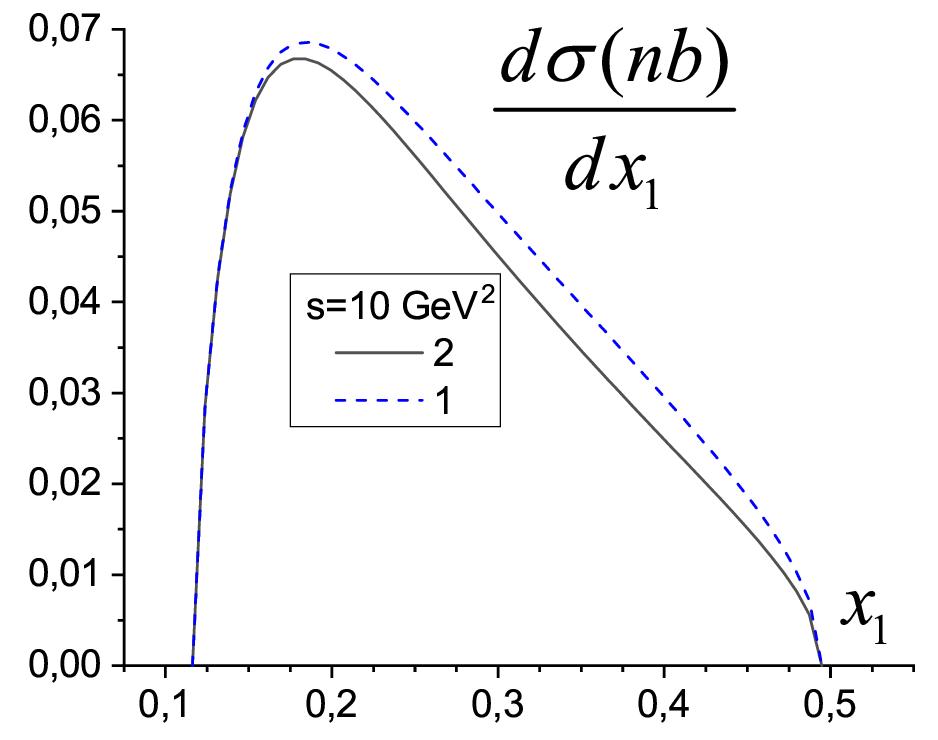}
\hspace{0.15cm}
\includegraphics[width=0.262\textwidth]{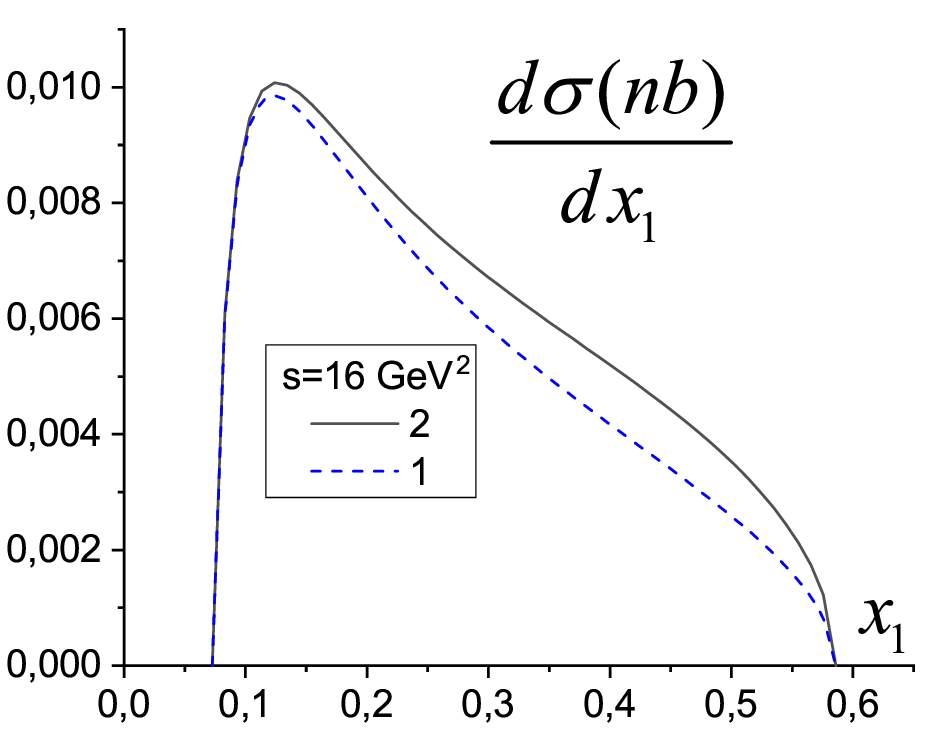}
\includegraphics[width=0.262\textwidth]{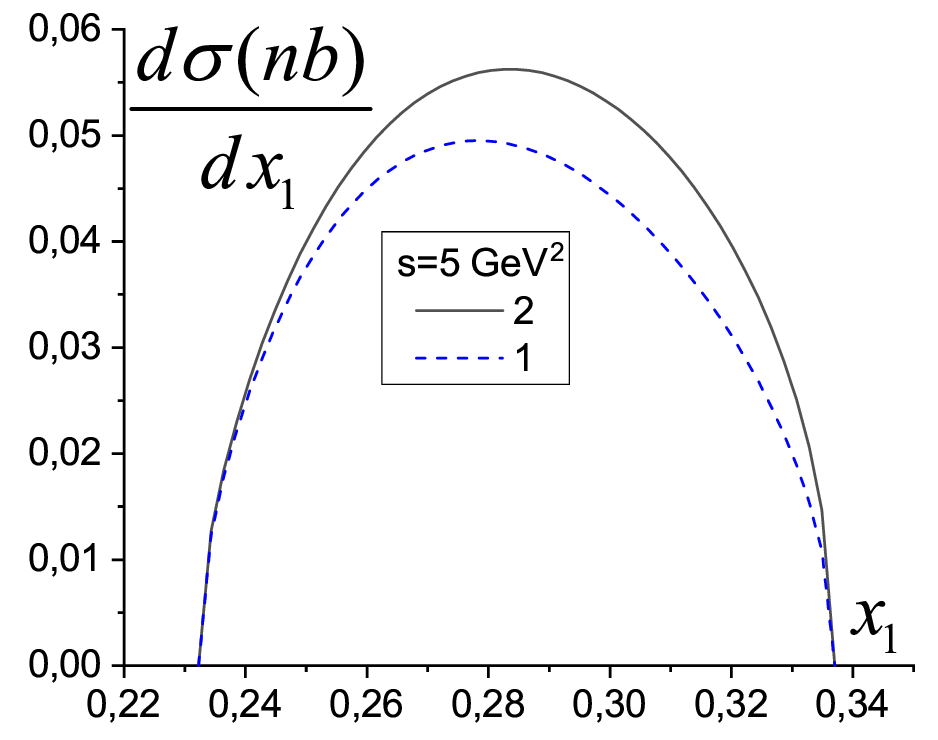}
\hspace{0.15cm}
\includegraphics[width=0.262\textwidth]{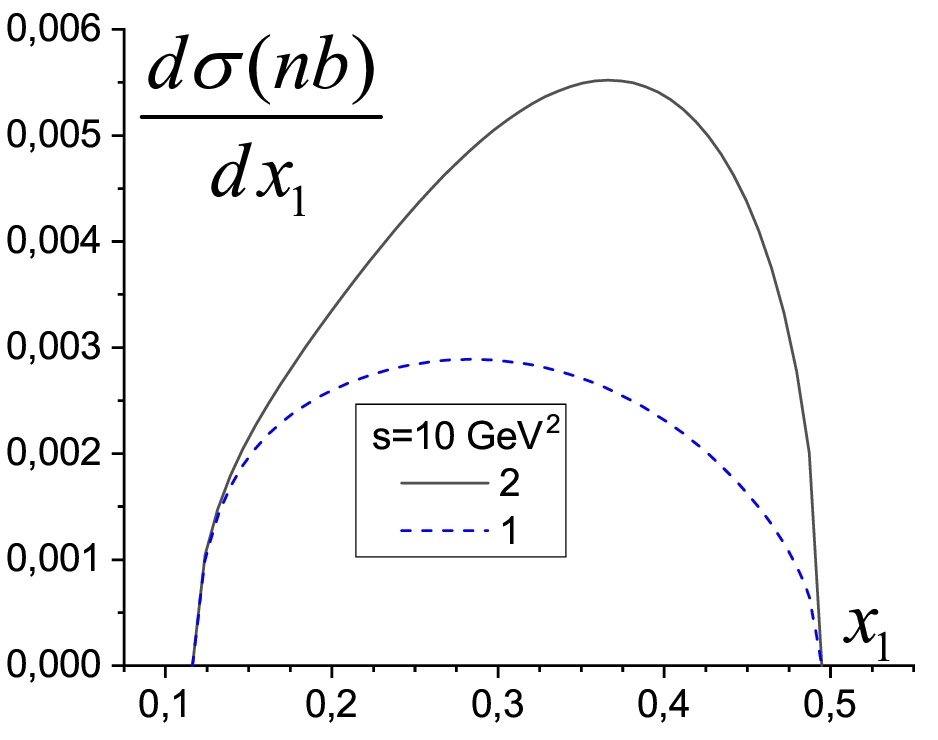}
\hspace{0.15cm}
\includegraphics[width=0.262\textwidth]{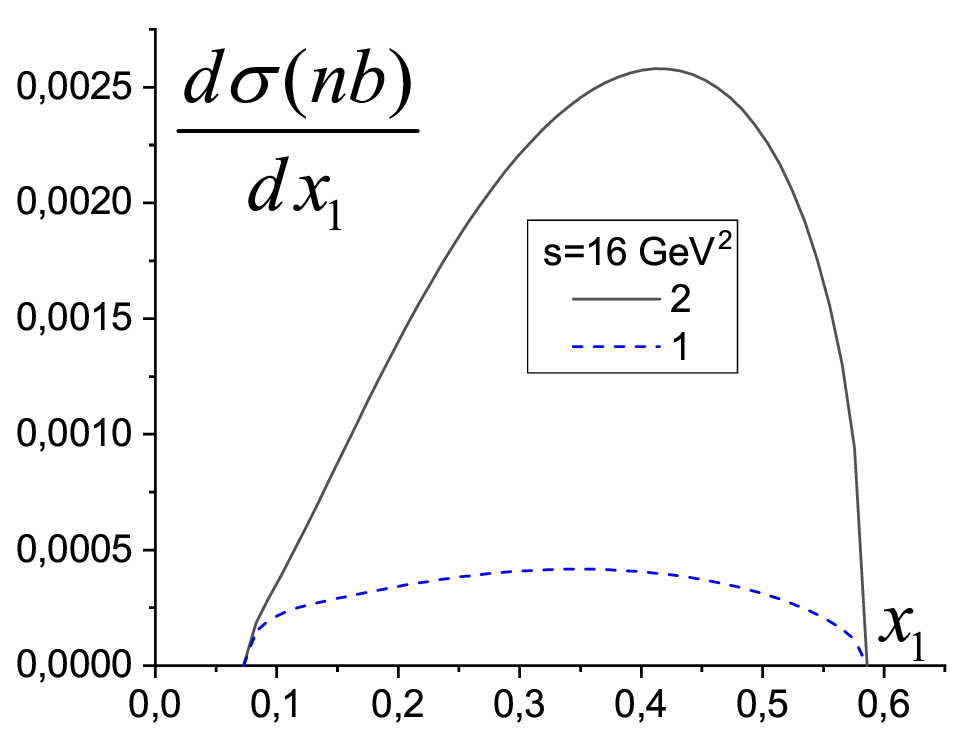}
\includegraphics[width=0.262\textwidth]{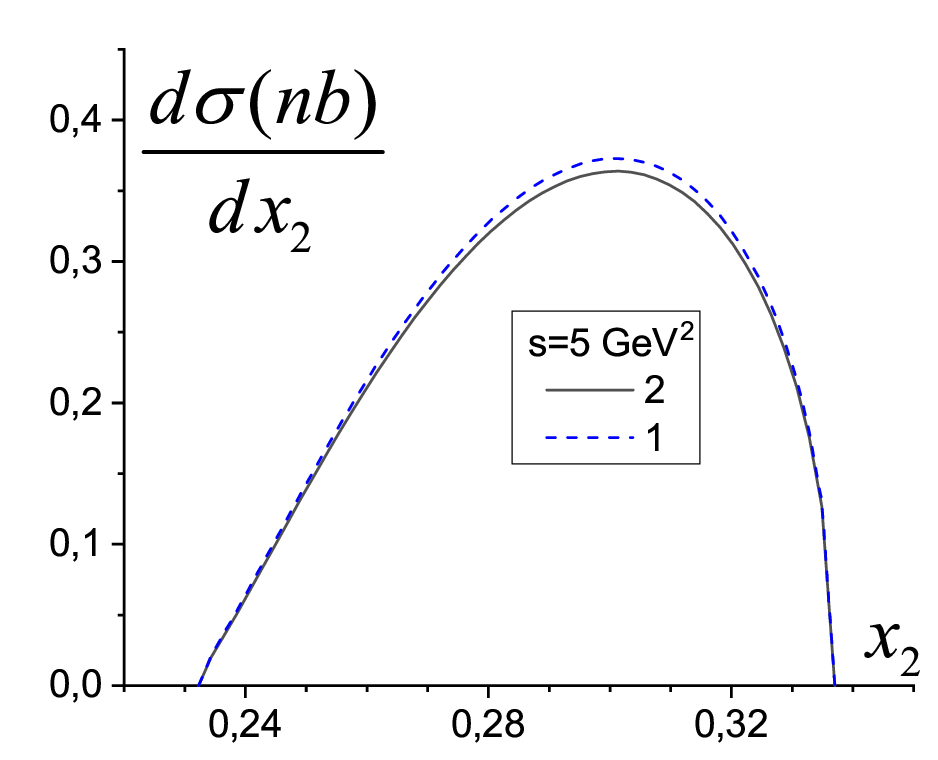}
\hspace{0.15cm}
\includegraphics[width=0.262\textwidth]{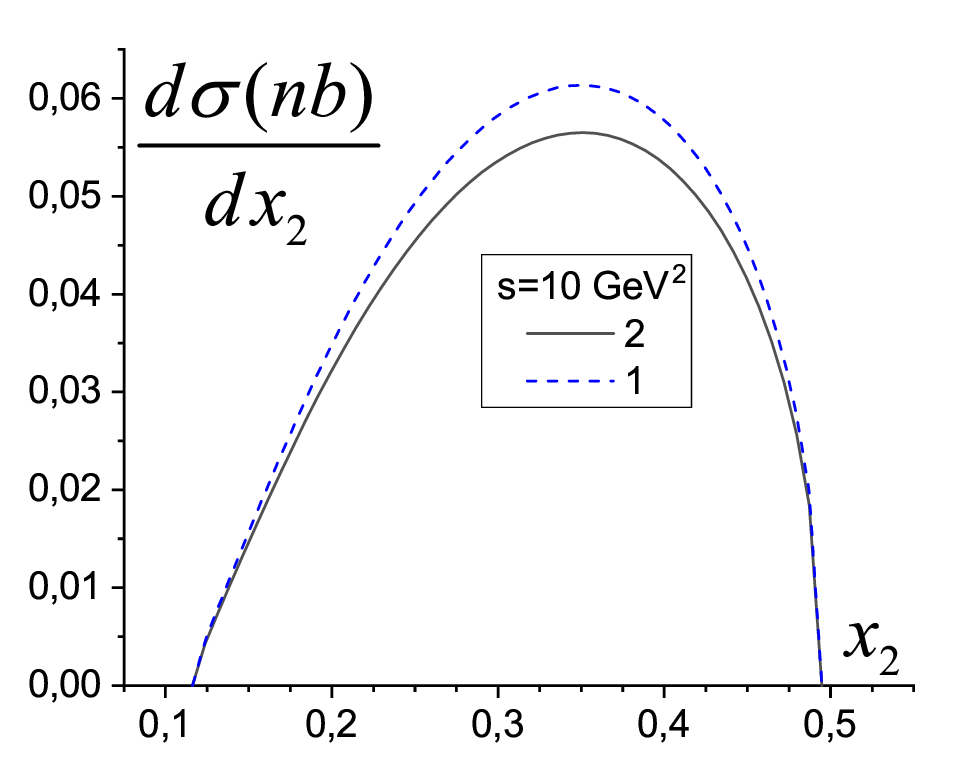}
\hspace{0.15cm}
\includegraphics[width=0.262\textwidth]{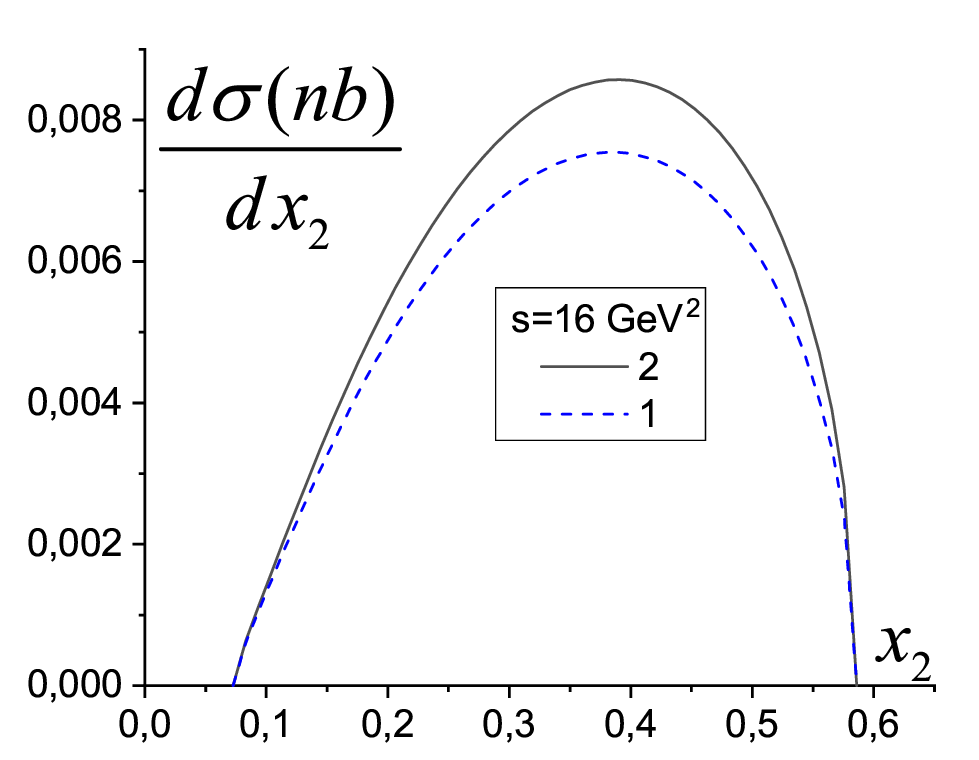}
\includegraphics[width=0.262\textwidth]{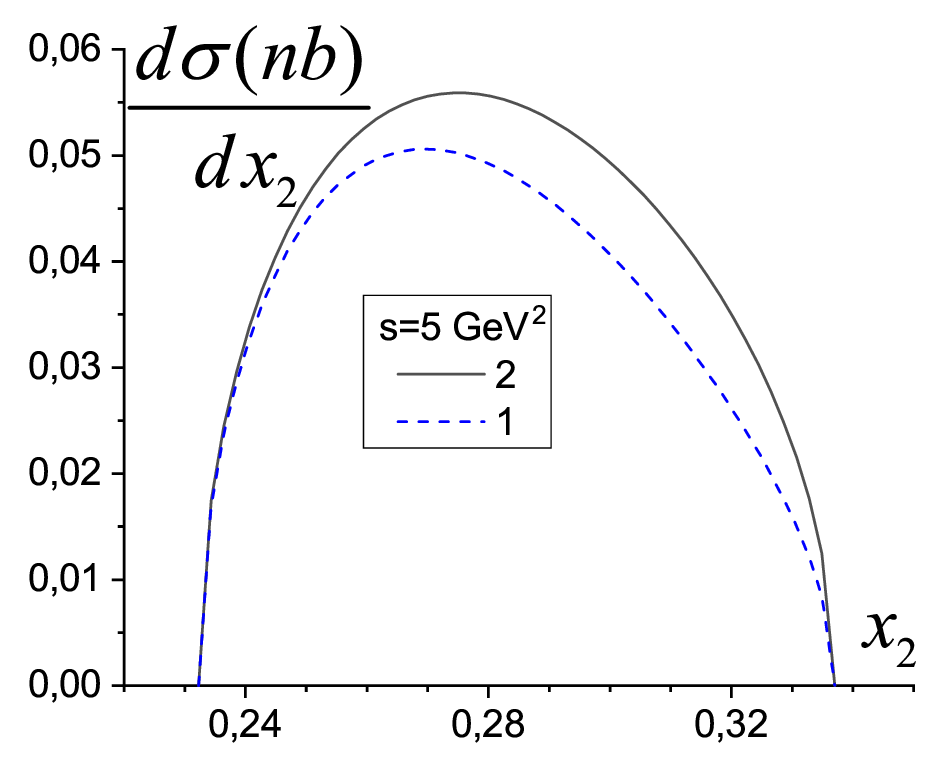}
\hspace{0.15cm}
\includegraphics[width=0.262\textwidth]{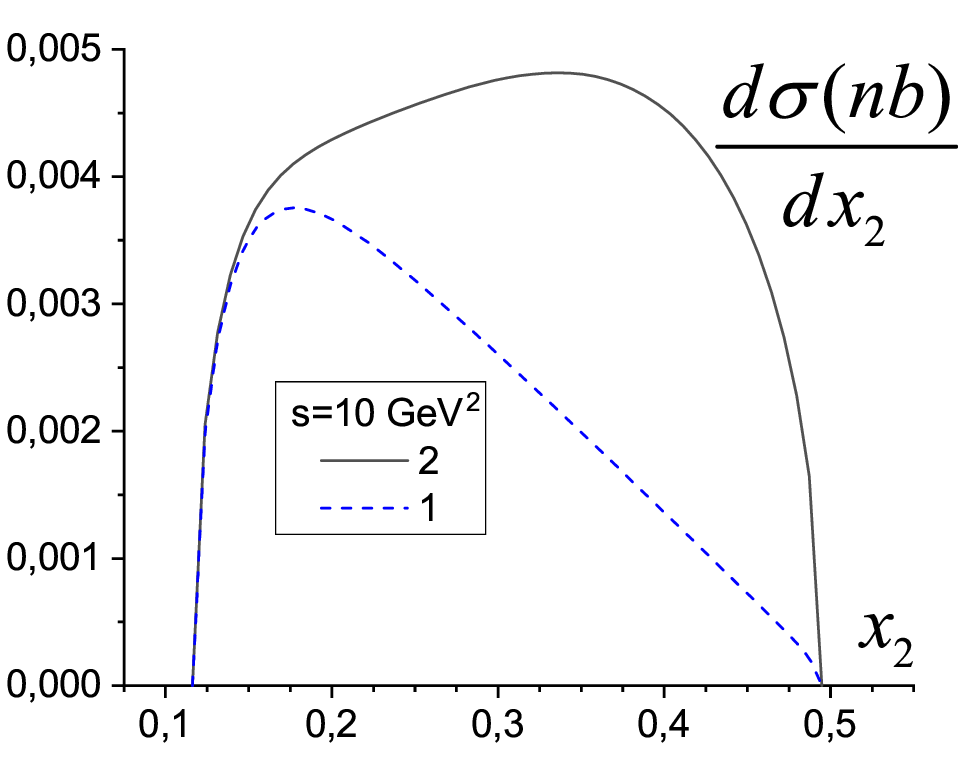}
\hspace{0.15cm}
\includegraphics[width=0.262\textwidth]{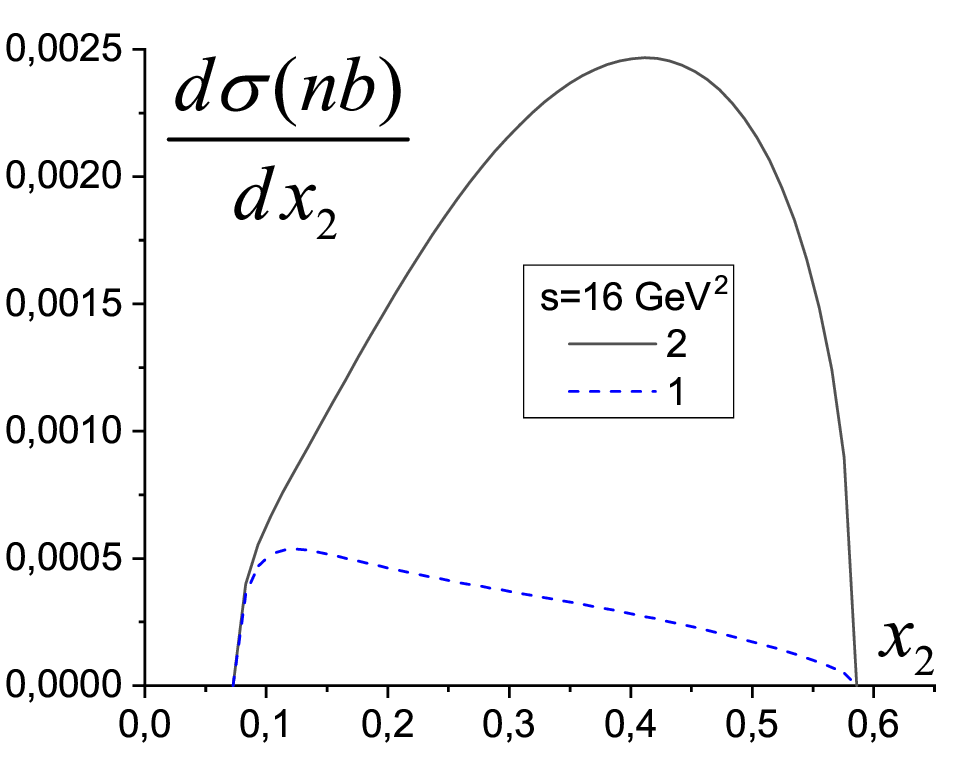}
\includegraphics[width=0.262\textwidth]{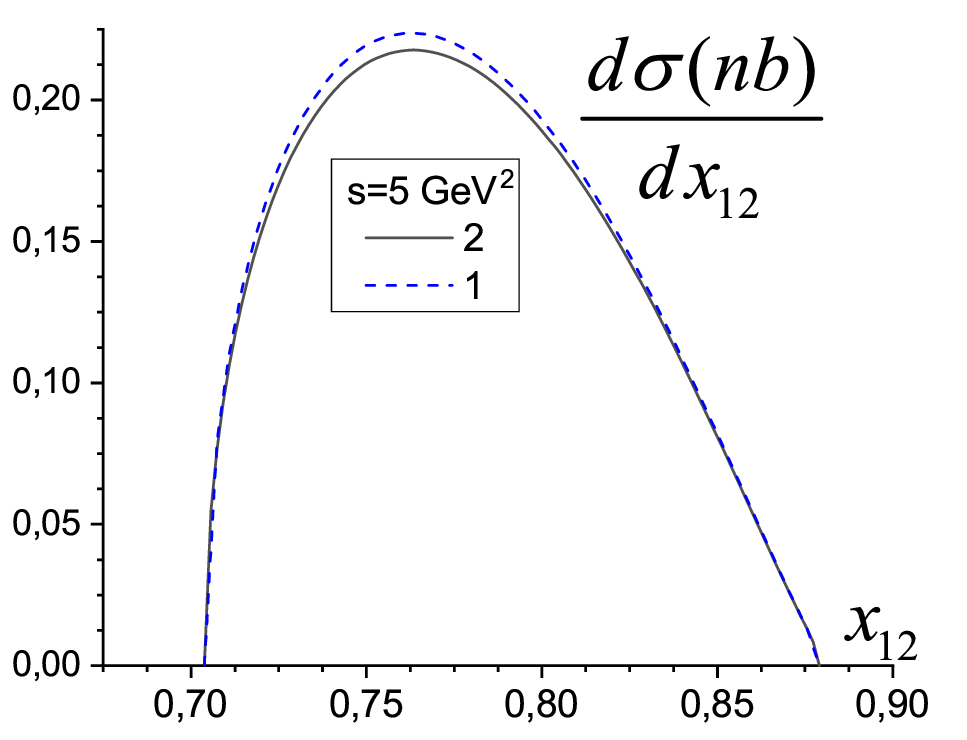}
\hspace{0.15cm}
\includegraphics[width=0.262\textwidth]{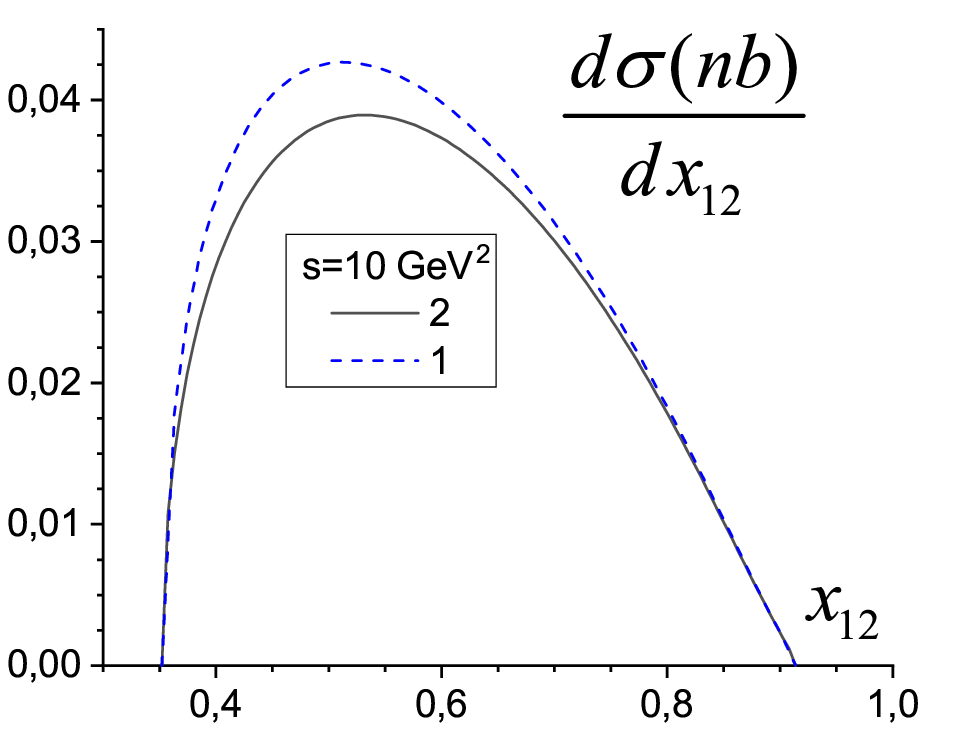}
\hspace{0.15cm}
\includegraphics[width=0.262\textwidth]{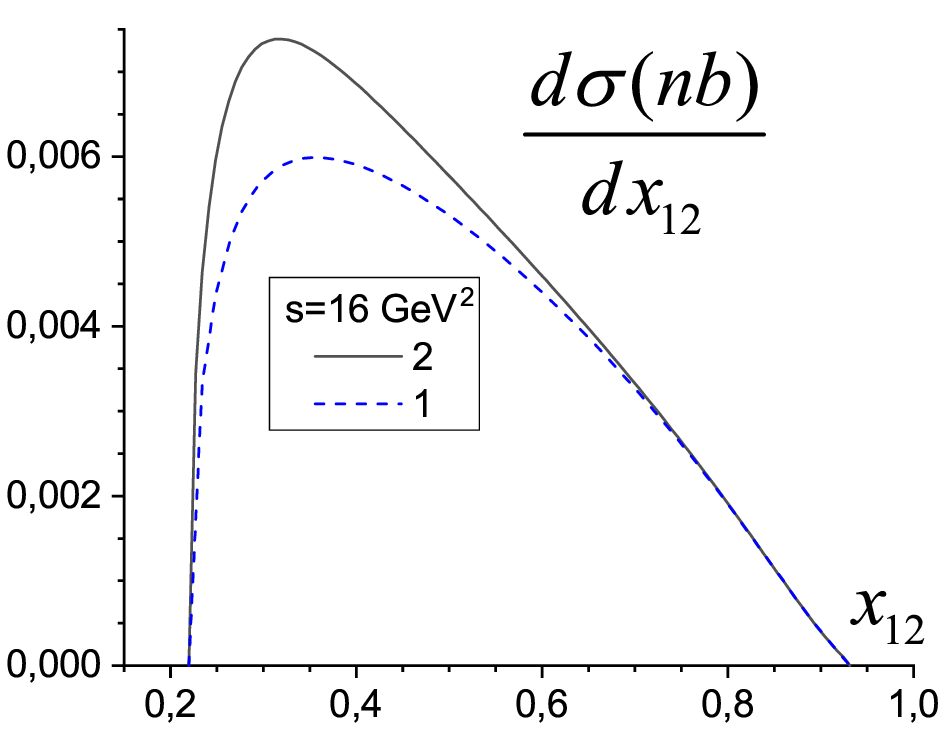}
\includegraphics[width=0.262\textwidth]{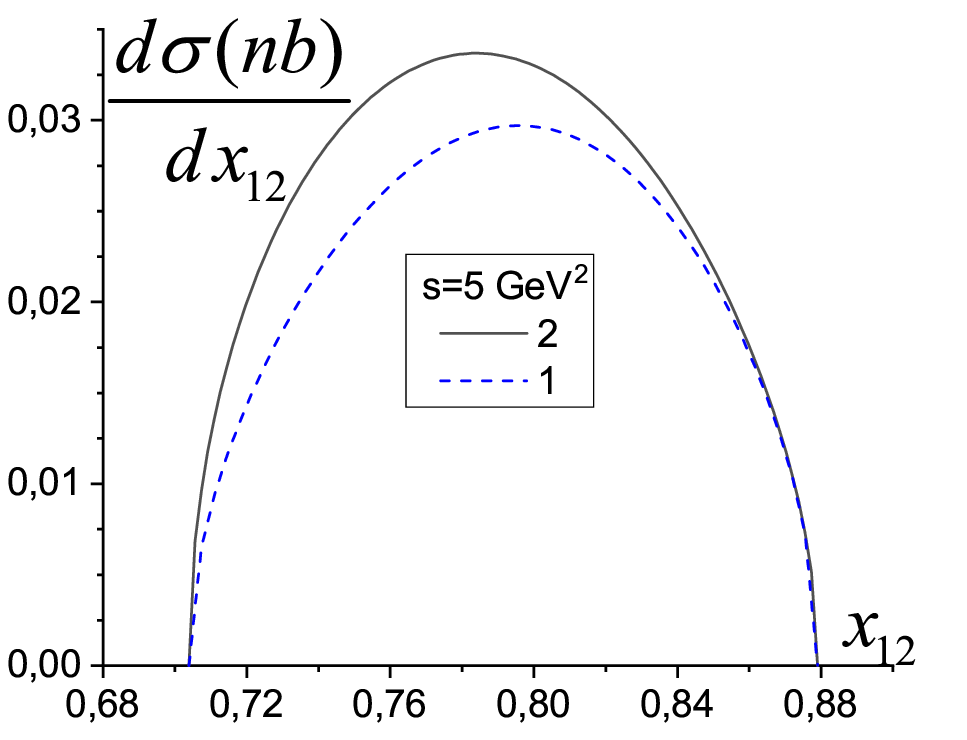}
\hspace{0.15cm}
\includegraphics[width=0.262\textwidth]{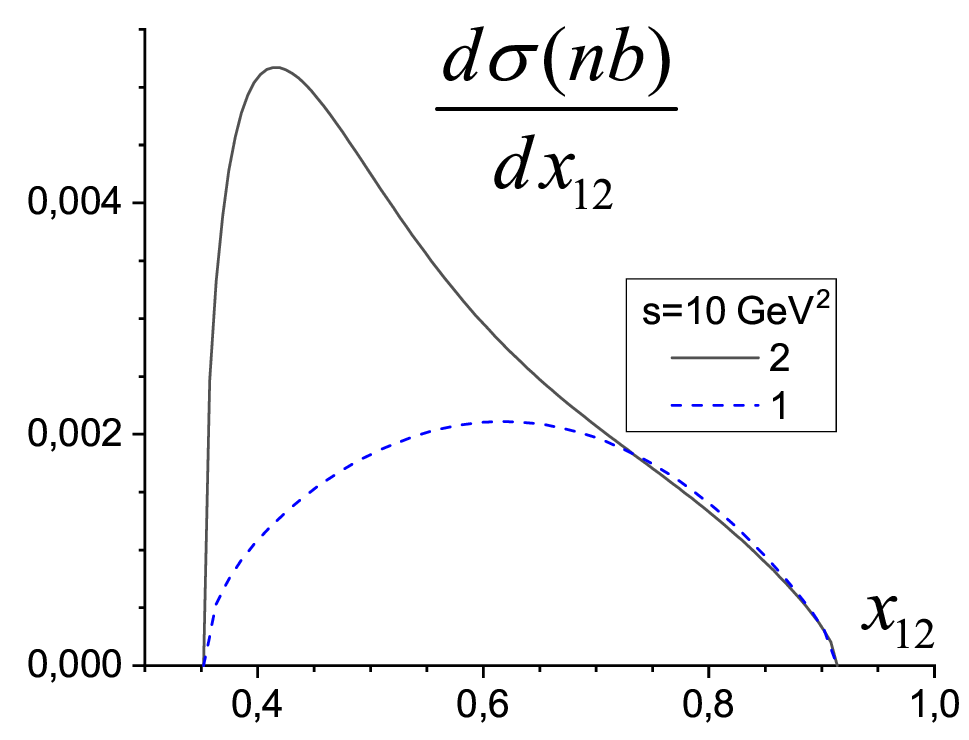}
\hspace{0.15cm}
\includegraphics[width=0.262\textwidth]{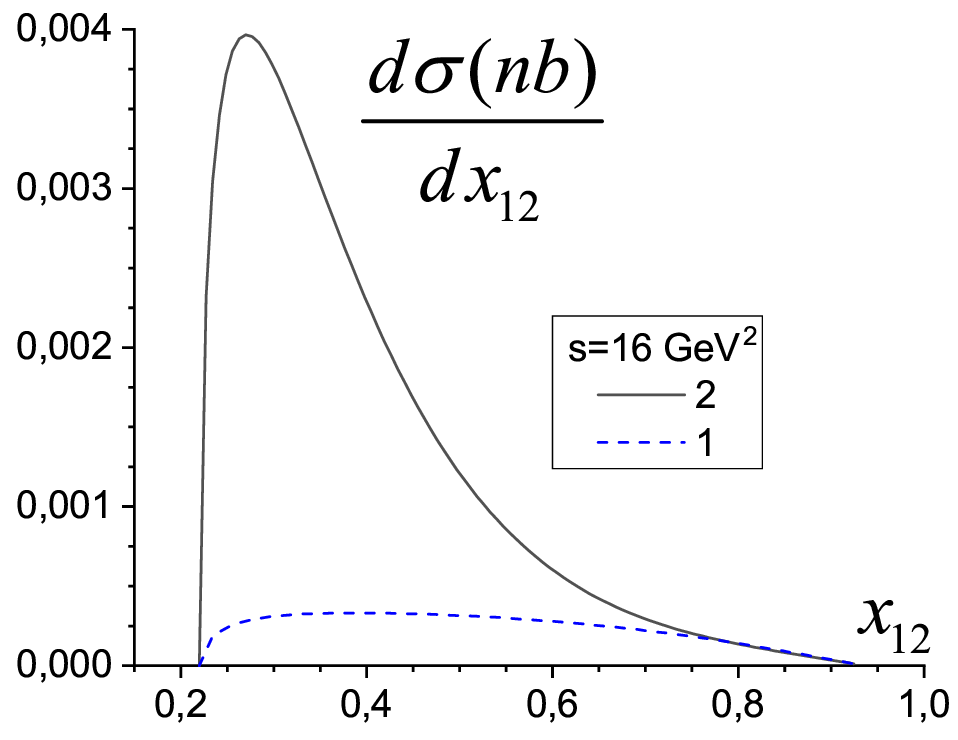}
 \parbox[t]{0.9\textwidth}
 {\caption{Single differential distributions over the dimensionless  variables $x_{1,2}=s_{1,2}/s$, $x_{12}=s_{12}/s $ in reaction  $e^+ + e^- \to p + \bar{n} +\pi^-$  at $s=5$, 10, 16 GeV$^2$.  The blue dotted curves labeled by "1" are obtained with $F_\pi =F_1^p-F_1^n$ and the black solid curves labeled by "2"  correspond to parametrization $F_\pi$ as given in Ref.\cite{BaBar:2012bdw}. The order of the "old" and "new" rows is the same as in Fig.\,2. }
 \label{Fig:Fig5}}
\end{figure}

\begin{figure}[h!]
\centering
\includegraphics[width=0.262\textwidth]{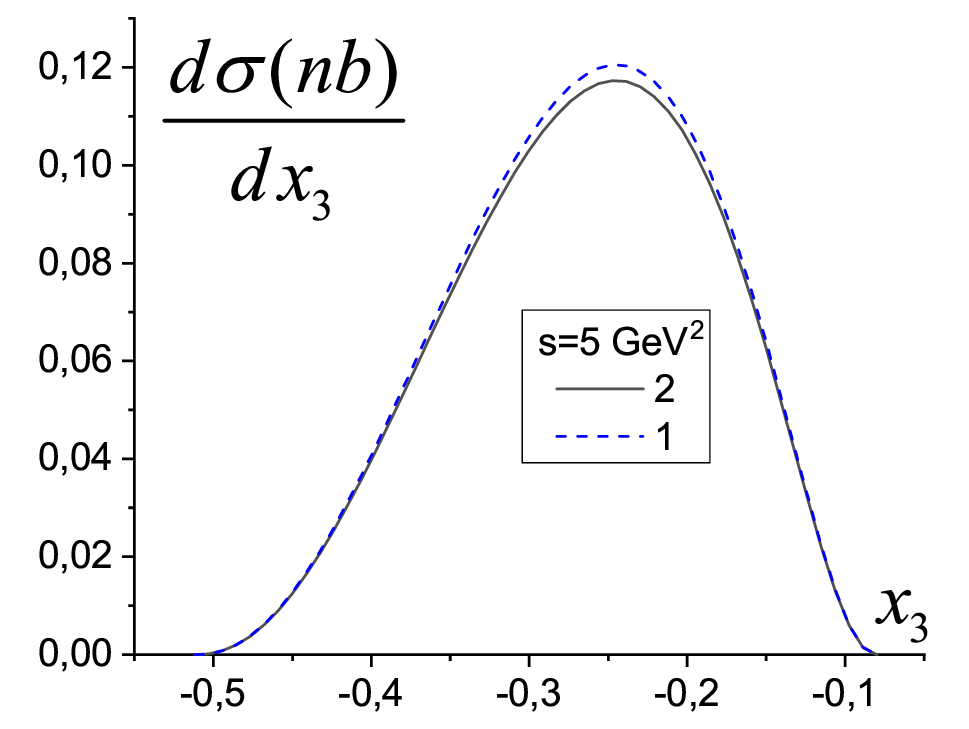}
\hspace{0.15cm}
\includegraphics[width=0.262\textwidth]{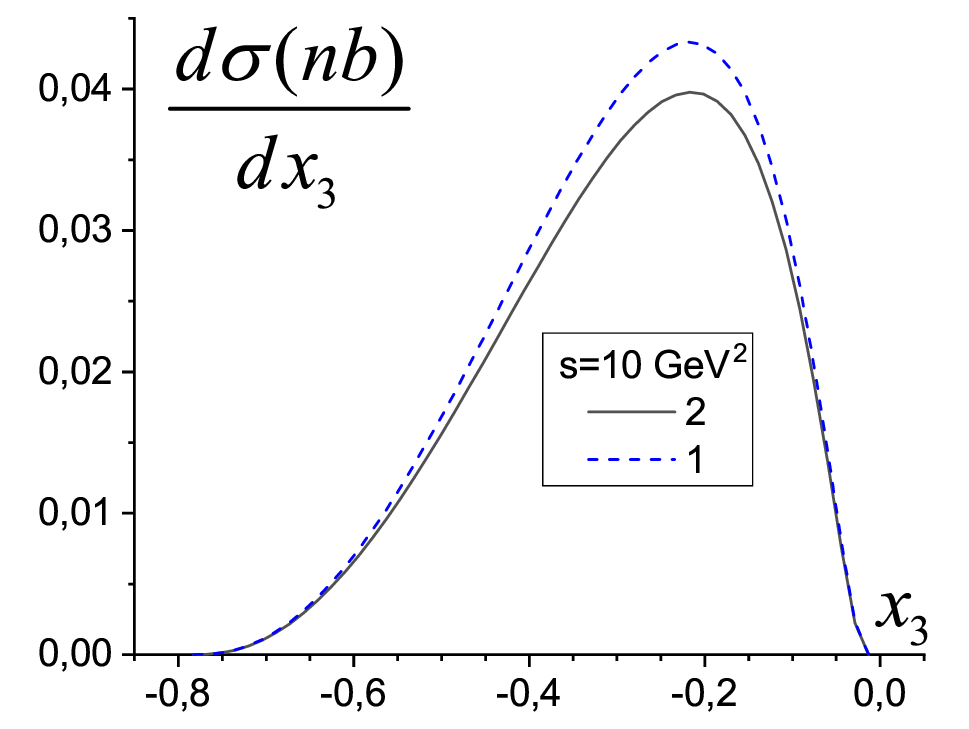}
\hspace{0.15cm}
\includegraphics[width=0.262\textwidth]{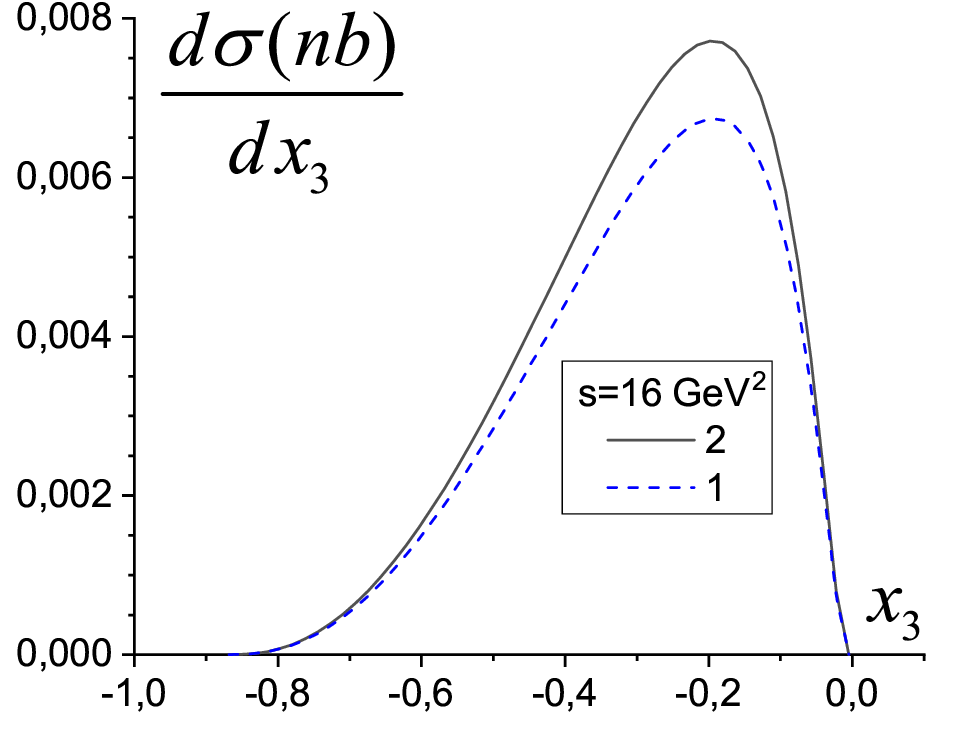}
\includegraphics[width=0.262\textwidth]{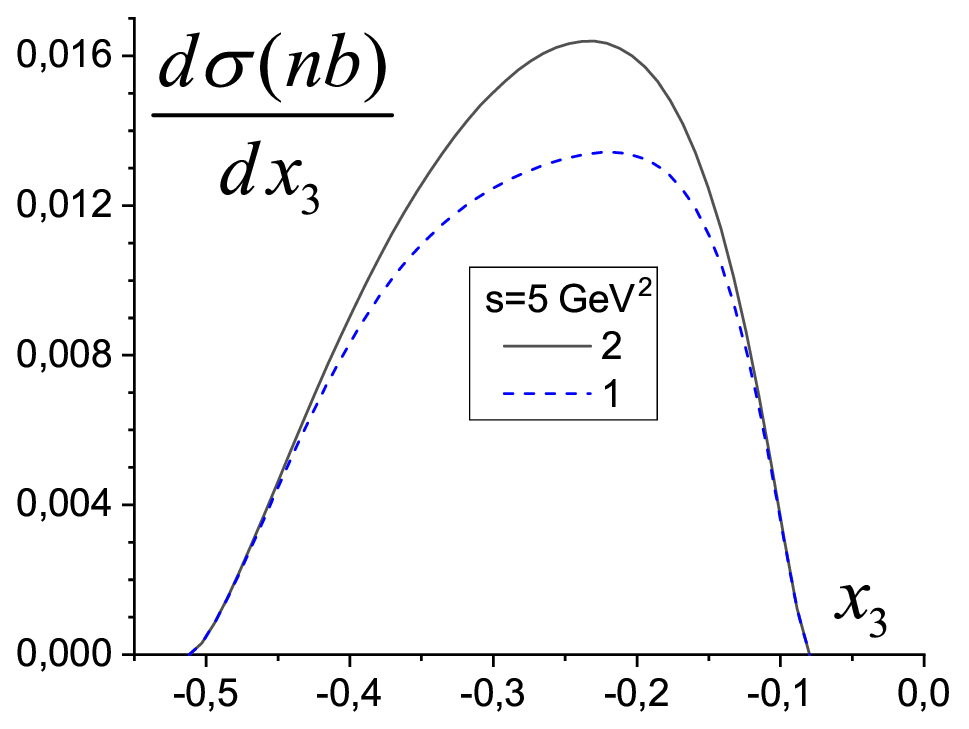}
\hspace{0.15cm}
\includegraphics[width=0.262\textwidth]{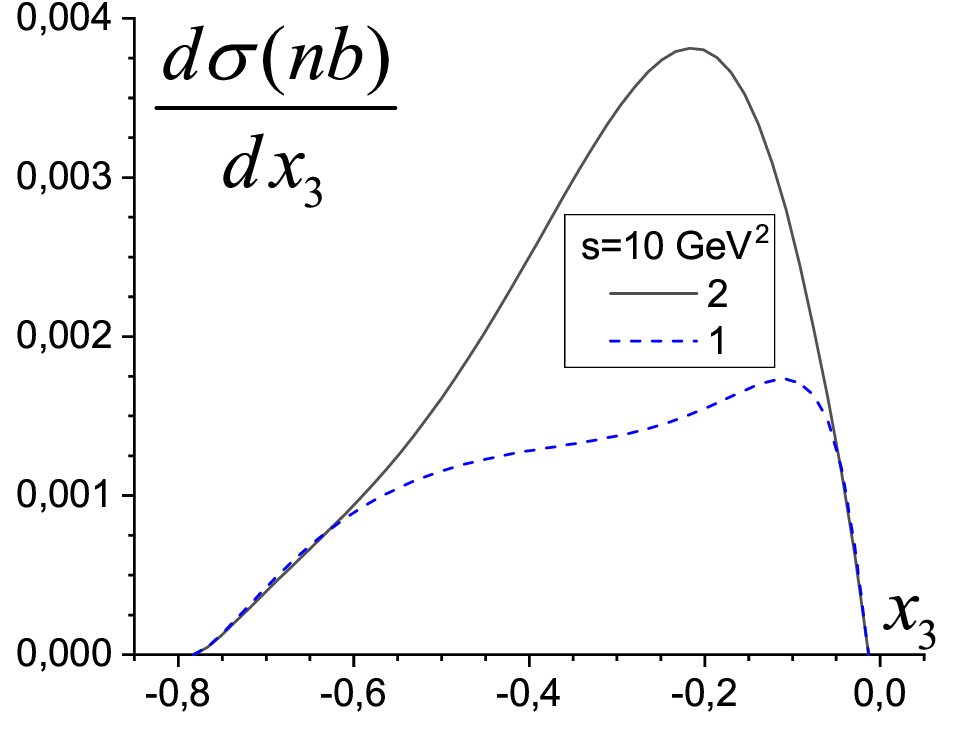}
\hspace{0.15cm}
\includegraphics[width=0.262\textwidth]{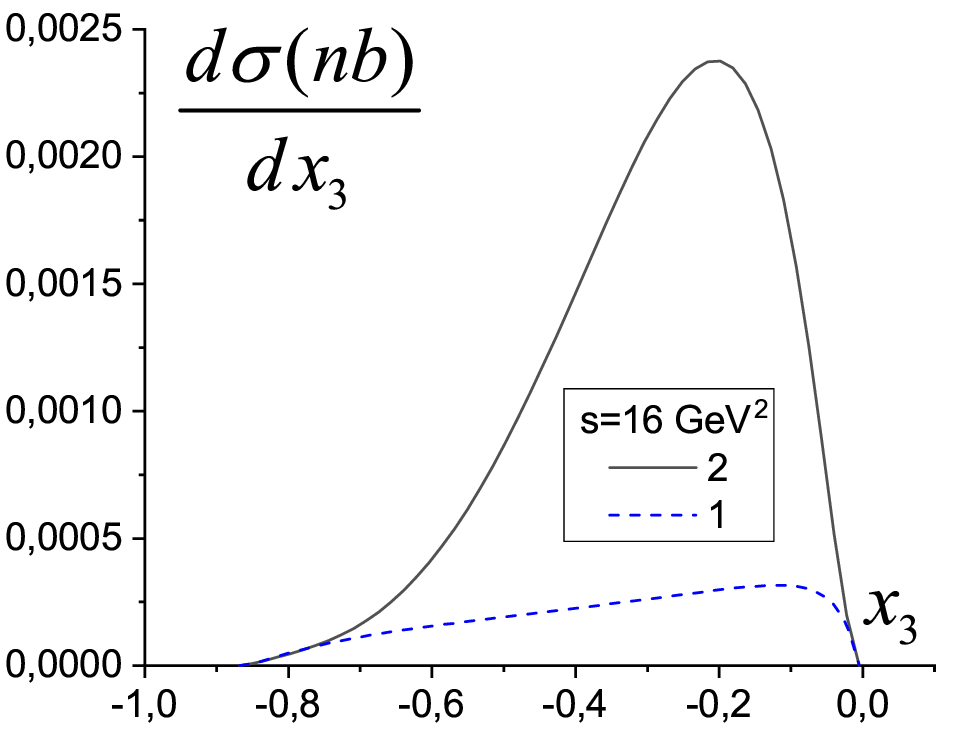}
\includegraphics[width=0.262\textwidth]{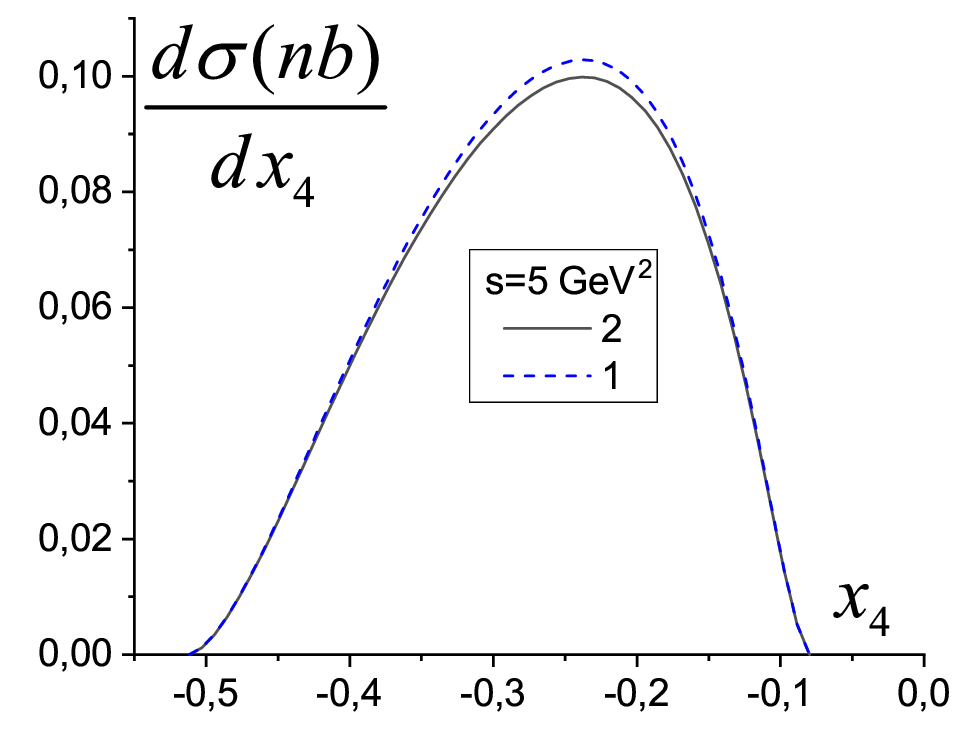}
\hspace{0.15cm}
\includegraphics[width=0.262\textwidth]{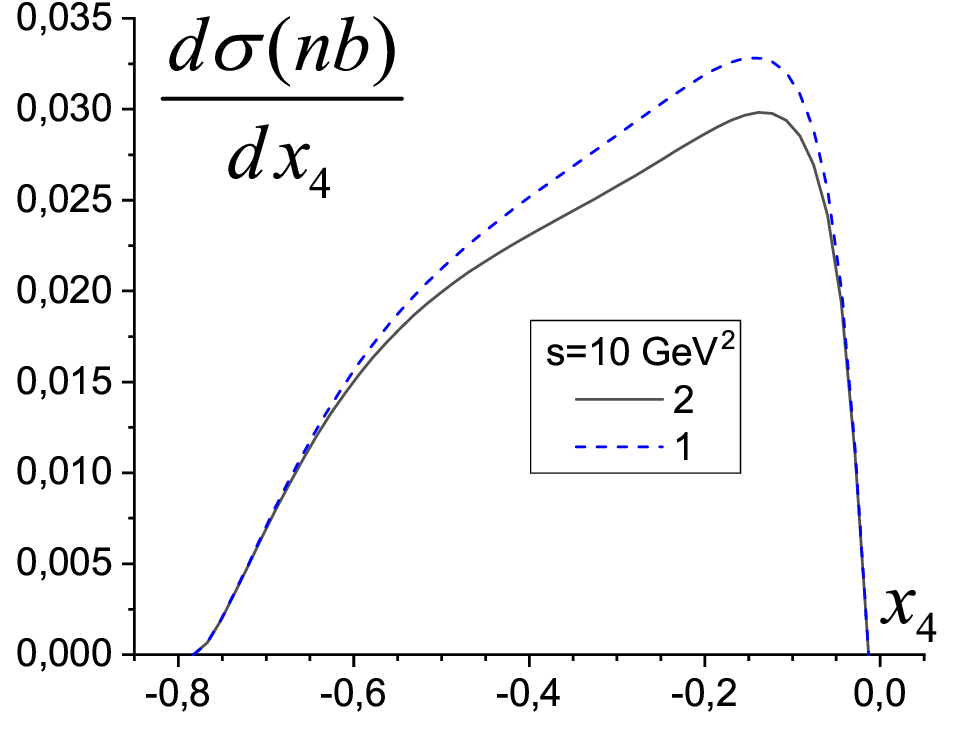}
\hspace{0.15cm}
\includegraphics[width=0.262\textwidth]{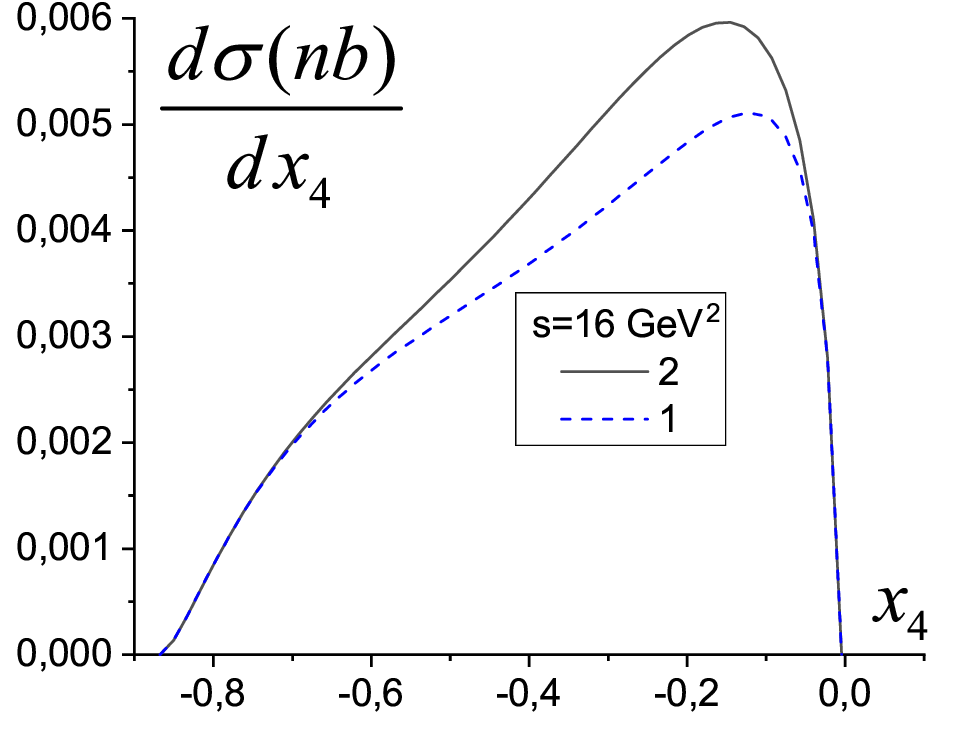}
\includegraphics[width=0.262\textwidth]{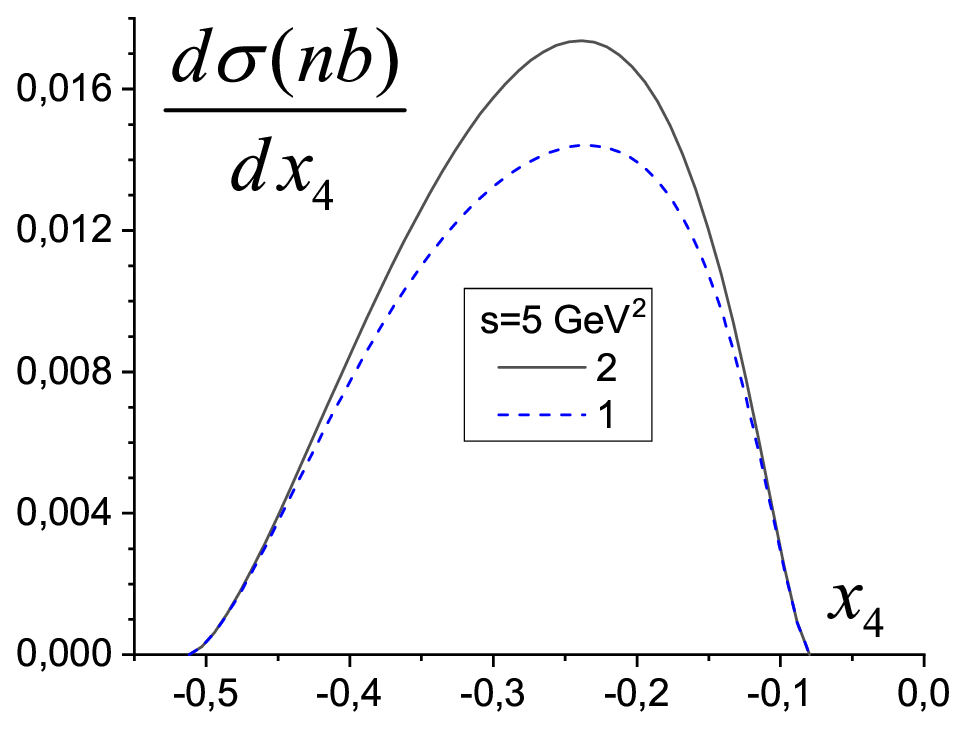}
\hspace{0.15cm}
\includegraphics[width=0.262\textwidth]{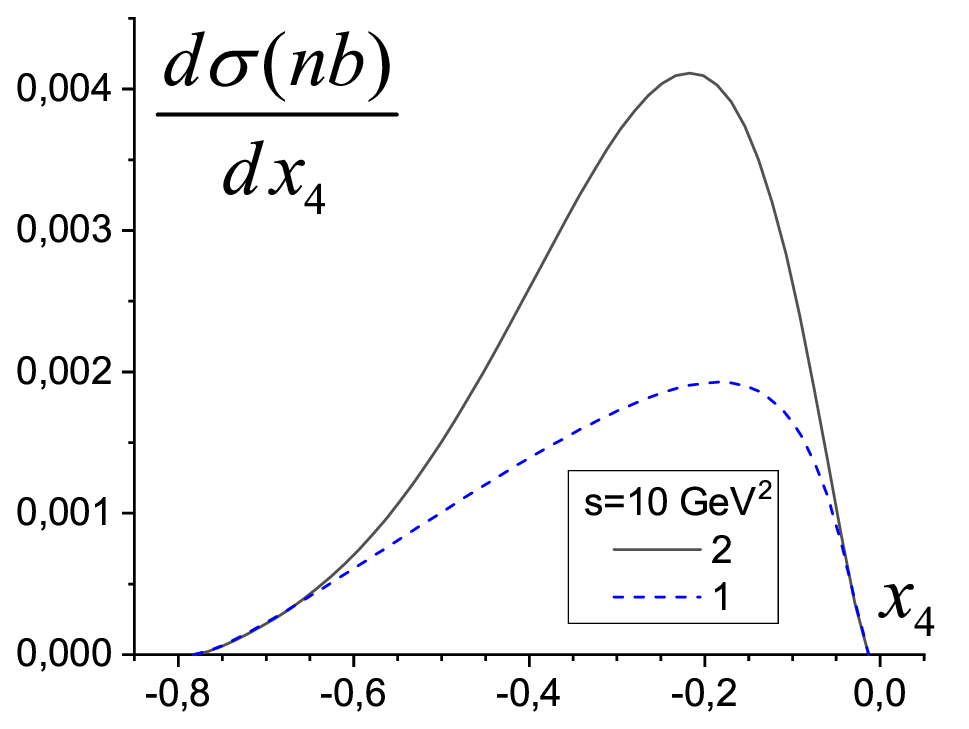}
\hspace{0.15cm}
\includegraphics[width=0.262\textwidth]{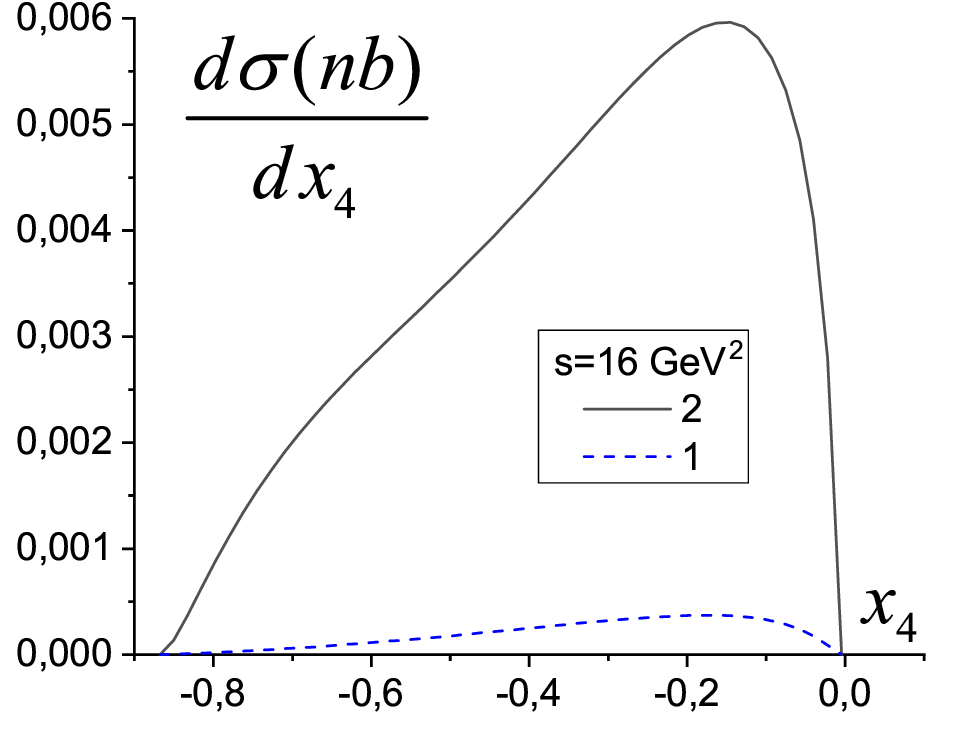}
\parbox[t]{0.9\textwidth}{\caption{The same as in Fig.\,5 but for distributions over the dimensionless  variables $x_{3,4}=t_{1,2}/s.$ .}
\label{Fig:Fig6}}
\end{figure}

It is evident that the $s_{12}$-distribution for the reaction $e^+ + e^- \to n + \bar{p} +\pi^+$ is the same, and the total cross sections in both channels (with $\pi^-$ and $\pi^+$) coincide. The single differential distributions over different dimensionless invariant variables are shown in Figs. \ref{Fig:Fig5}, \ref{Fig:Fig6}.

The distributions over $x_1\,(x_3)$ for the reactions $e^+ + e^- \to n + \bar{p} +\pi^+$ coincides with the distributions over $x_2\,(x_4)$ for reaction
$e^+ + e^- \to p + \bar{n} +\pi^-.$

\begin{figure}[h!]
\centering
\includegraphics[width=0.42\textwidth]{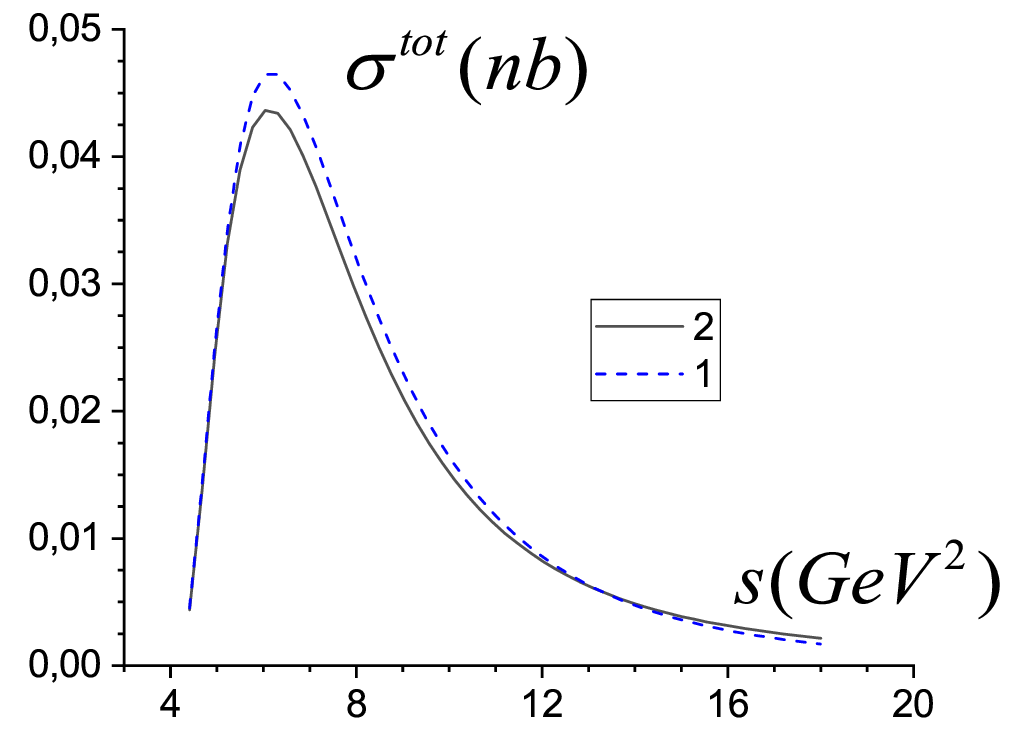}
\includegraphics[width=0.42\textwidth]{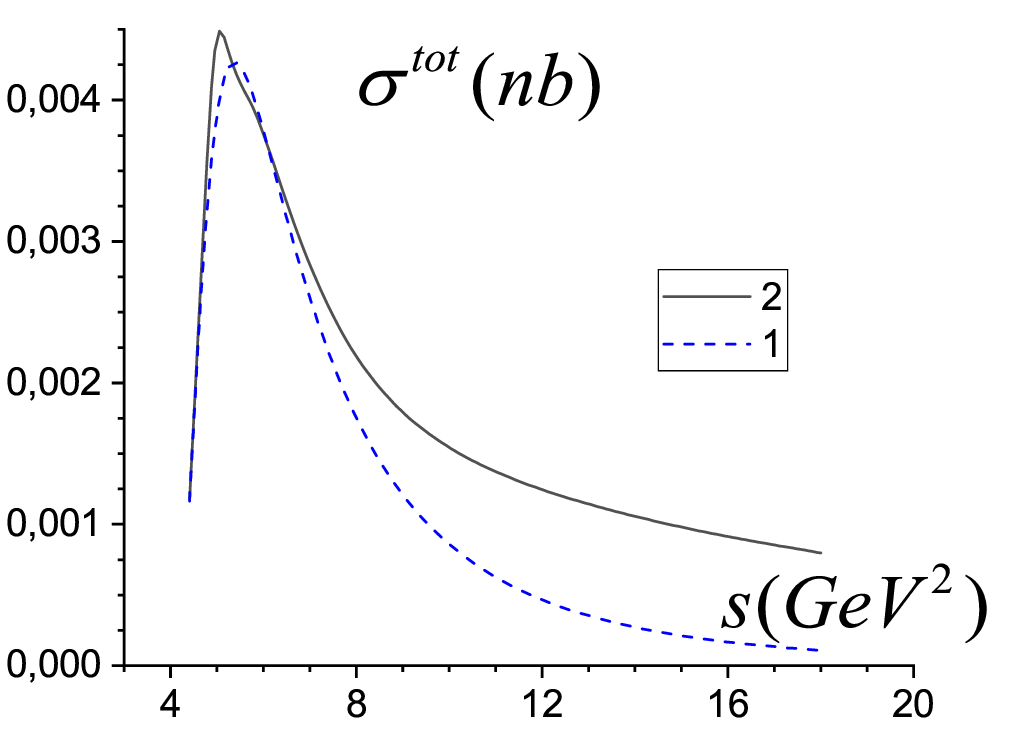}
\includegraphics[width=0.42\textwidth]{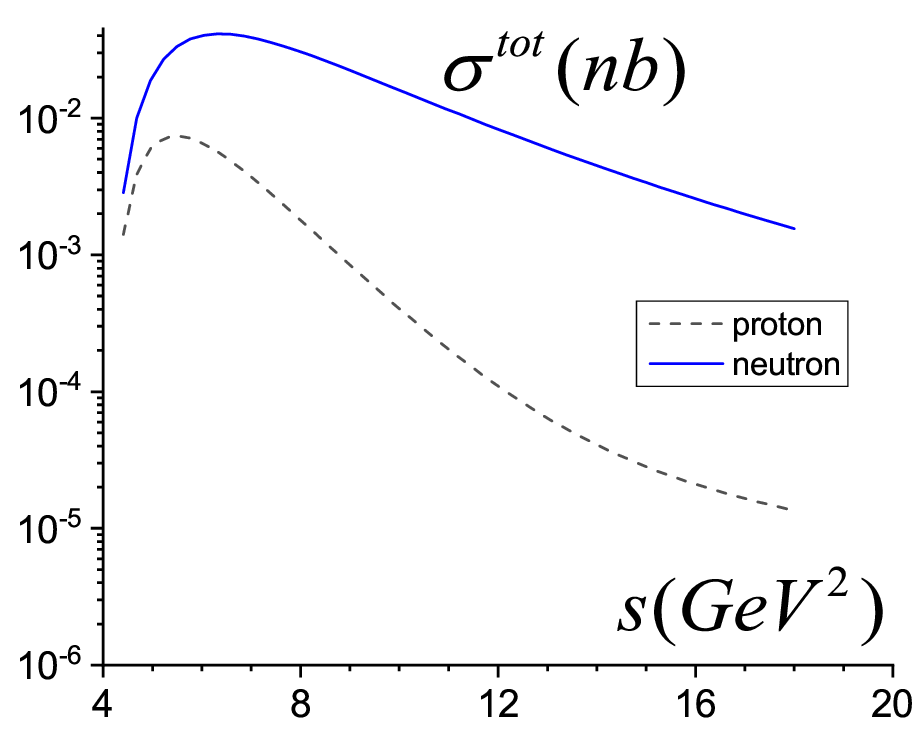}
\includegraphics[width=0.42\textwidth]{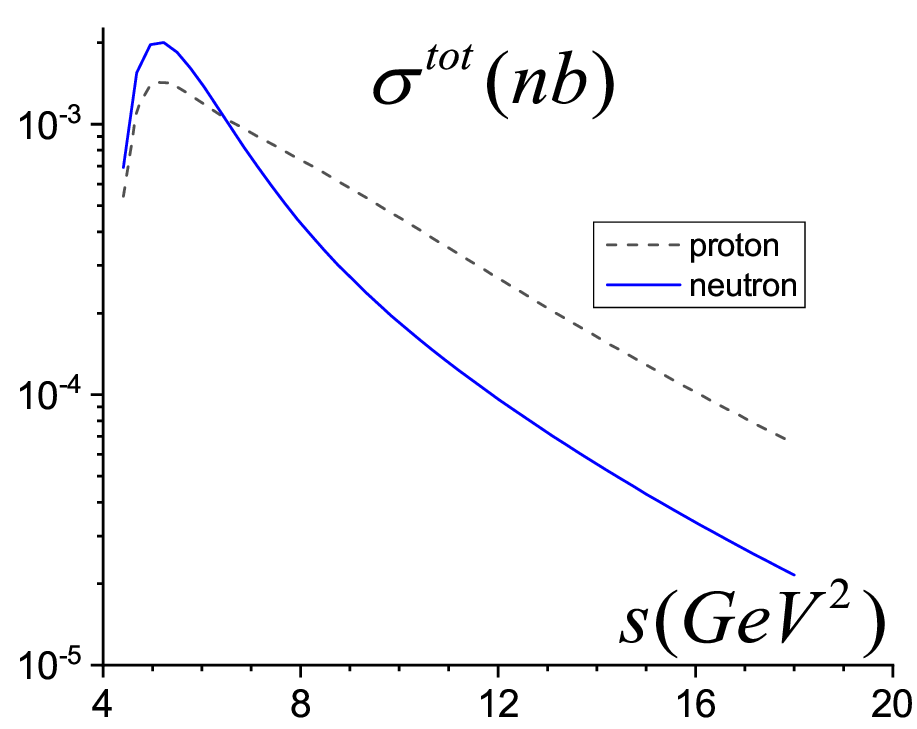}
 \parbox[t]{0.9\textwidth}{\caption{The upper row: total cross section of the reaction  $e^+ + e^- \to p + \bar{n} +\pi^-;$ the left (right) panel corresponds to the "old" ("new") parametrization of the nucleon form factors. In the lower row we plotted the total cross section for the $\pi^0 N \bar{N}$,  $N=p,\,n$ channels with the same parameterizations of the nucleon form factors.}\label{Fig:Fig7}}
\end{figure}

In Fig. \ref{Fig:Fig7} the total cross section of the reaction the $e^+ + e^- \to p + \bar{n} +\pi^-$ is drawn for the two different parameterizations of the nucleon and pion electromagnetic form factors. For comparison we also plot the total cross section of processes with the neutral pion $e^+ + e^- \to N + \bar{N} +\pi^0, \ N=p,\,n,$ (where the diagram  of the intermediate pion is absent) with the same parameterizations of the nucleon form factors.

\section{Discussion and Conclusions}
\label{Section:discussion}
In this article we studied the contribution of the nonresonant mechanism to the reactions $e^+ + e^- \to p + \bar{n} +\pi^-$ and $e^+ + e^- \to n + \bar{p} +\pi^+$.  The double and single distributions over invariant variables as well the total cross sections were calculated.
The formalism developed in our previous work \cite{Gakh:2022fad} was followed for the calculation of the matrix element over the invariant amplitudes and of the final particles phase space, to derive the matrix element squared and to perform the analytical integrations. To apply this formalism we modify the hadron current (to make it conserved) by well known procedure, see Eq.\,(\ref{eq:I3}), \cite{Akhiezer:1977nd} that does not change the matrix element but allows to expend
 definitely the hadron tensor with the invariant amplitudes Eq.\,(\ref{eq:invamp1}).

 First we expressed the invariant amplitudes $A_i$ in terms of proton, neutron and pion FF's and then obtained the hadronic tensor using the results of our paper \cite{Gakh:2022fad}. We calculated the matrix element squared that is the convolution of the hadron and lepton tensors and made sure at this level that the substitution $F_{1,2}^p(q^2)\rightleftarrows F_{1,2}^n(q^2),\, F_(q^2)\pi\rightarrow - F_\pi(q^2) $ is equivalent to the change $s_1\rightleftarrows s_2; t_1\rightleftarrows t_2.$
Because the phase space of the final hadrons is invariant under this last change, this allows to understand the form of different distributions for the reaction $e^+ + e^- \to \pi^+ + n +\bar{p},$\, looking at plots in Figs.\,\ref{Fig:Fig3}$-$\ref{Fig:Fig6}. For example, the distribution over the dimensionless variables $[x_1\,(x_2),\,x_{12}]$ in Fig.\,\ref{Fig:Fig3} is the same as distribution over $[x_2\,(x_1),\,x_{12}]$ for the reaction with positive charged pion. In Fig.\,\ref{Fig:Fig4} the distribution over $[x_1\,(x_2), x_4\,(x_3)]$ is the same as the $[x_2\,(x_1), x_3\,(x_4)]$ one in the case of $\pi^+$ production. As concerns the single distributions, it is obvious that the $(s_{12})$ one in Fig.\,\ref{Fig:Fig5} is identical for both channels whereas the $[x_1\,(x_2)]$ and $[x_3\,(x_4)]$ plots in Figs.\,\ref{Fig:Fig5}, \ref{Fig:Fig6} correspond to $[x_2\,(x_1)]$ and $[x_4\,(x_3)]$ plots for the $\pi^+$-channel. The total cross sections plotted in Fig.\,\ref{Fig:Fig7} are equal for both channel.

Formulas for different double and single distributions (except the double $(t_1,\,t_2),$ single
$t_1$ and $t_2$ ons) were derived analytically and checked within the Mathematica framework \cite{Math:2024} that was also used for plotting. The rest distributions as well the total cross sections were obtained by one numerical integration.

In our calculations we used two different parameterizations of the nucleon form factors (labeled as "old"\cite{Iachello:1972nu,Iachello:2004aq} and "new" \cite{Bijker:2004yu} in \cite{Gakh:2022fad}) as well of the pion one. One parametrization \cite{BaBar:2012bdw} corresponds to the  monopole behaviour of $F_\pi(q^2)$ at large $q^2$ in accordance with the quark counting rule and another is defined by the difference of the proton and neutron Dirac
electromagnetic form factors, see Eq.(\ref{eq:FF1pi}). The last one is against the quark counting rule but we used it because in this case the conservation of the hadron current does not require any modification.




\end{document}